\documentclass[aps, prb, superscriptaddress,  floatfix, reprint]{revtex4-1}
\newcommand{\rem}[1]{} 
\newcommand{\RM}[1]{\MakeUppercase{\romannumeral #1}}
\newcommand{\TB}[0]{\ensuremath{T}} 
\newcommand{\RW}[0]{\ensuremath{R_\mathrm{NW}}} 
\newcommand{\VS}[0]{\ensuremath{V_\mathrm{S}}} 
\newcommand{\lw}[0]{\ensuremath{l_\mathrm{NW}}} 
\newcommand{\Lambdael}[1]{\ensuremath{\Lambda_\mathrm{el#1}}} 
\newcommand{\Lambdath}[1]{\ensuremath{\Lambda_\mathrm{th#1}}} 
\newcommand{\RT}[1]{\ensuremath{\mathrm{RT}}} 
\newcommand{\Reph}[1]{\ensuremath{R_\mathrm{e-ph}}} 
\newcommand{\Tc}[1]{\ensuremath{T_\mathrm{\RM{1}-\RM{2}}}}

\usepackage{ifpdf}
\usepackage{amssymb}
\usepackage{MnSymbol}
\usepackage{siunitx}
\DeclareSIUnit\rpm{rpm} 
\DeclareSIUnit{\molar}{M}
\usepackage[version=3]{mhchem}
\usepackage{graphicx}

\usepackage[usenames,dvipsnames,svgnames,table]{xcolor}
\usepackage{epstopdf}

\begin{document}
\title{Temperature-Dependent Thermoelectric Properties of Individual Silver Nanowires}

\author{D. Kojda}
\affiliation{AG Neue Materialien, Humboldt-Universit\"at zu Berlin, 10099 Berlin, Germany}
\author{R. Mitdank}
\affiliation{AG Neue Materialien, Humboldt-Universit\"at zu Berlin, 10099 Berlin, Germany}
\author{M. Handwerg}
\affiliation{AG Neue Materialien, Humboldt-Universit\"at zu Berlin, 10099 Berlin, Germany}
\author{A. Mogilatenko}
\affiliation{Ferdinand-Braun-Institut, Leibniz-Institut f\"ur H\"ochstfrequenztechnik, 12489 Berlin, Germany}
\author{Z.~Wang}
\affiliation{Laboratory for Design of Microsystems, University of Freiburg -- IMTEK, 79110 Freiburg, Germany}
\author{J. Ruhhammer}
\affiliation{Laboratory for Design of Microsystems, University of Freiburg -- IMTEK, 79110 Freiburg, Germany}
\author{M. Kr\"oner}
\affiliation{Laboratory for Design of Microsystems, University of Freiburg -- IMTEK, 79110 Freiburg, Germany}
\author{P. Woias}
\affiliation{Laboratory for Design of Microsystems, University of Freiburg -- IMTEK, 79110 Freiburg, Germany}
\author{M. Albrecht}
\affiliation{Leibniz-Institut f\"ur Kristallz\"uchtung -- IKZ, 12489 Berlin, Germany}
\author{S. F. Fischer}
\affiliation{AG Neue Materialien, Humboldt-Universit\"at zu Berlin, 10099 Berlin, Germany}

\date{\today}

\begin{abstract}
Individual highly pure single crystalline silver nanowires (Ag NWs) were investigated with regard to the electrical conductivity $\sigma$, the thermal conductivity $\lambda$ and the Seebeck coefficient $S$ as function of the temperature $T$ between \SI{1.4}{\kelvin} and room temperature (RT). Transmission electron microscopy was performed subsequently to the thermoelectric characterization of the Ag NWs, so that their transport properties can be correlated with the structural data. The crystal structure, surface morphology and the rare occurrence of kinks and twinning were identified. The thermoelectric properties of the Ag NWs are discussed in comparison to the bulk: $S_{\mathrm{Ag,Pt}}(T)$ was measured with respect to platinum and is in agreement with the bulk, $\sigma(T)$ and $\lambda(T)$ showed reduced values with respect to the bulk. The latter are both notably dominated by surface scattering caused by an increased surface-to-volume ratio. By lowering $T$ the electron mean free path strongly exceeds the NW's diameter of \SI{150}{\nano\meter} so that the transition from diffusive transport to quasi ballistic one dimensional transport is observed. An important result of this work is that the Lorenz number $L(T)$ turns out to be independent of surface scattering. Instead the characteristic of $L(T)$ is determined by the material's purity. Moreover, $\sigma(T)$ and $L(T)$ can be described by the bulk Debye temperature of silver. A detailed discussion of the temperature dependence of $L(T)$ and the scattering mechanisms is given.
\end{abstract}

\pacs{65.80.-g, 72.15.Jf, 73.63.-b, 81.07.Gf}


\maketitle 

\section{Introduction}
\label{into}
Extensive efforts have been done to produce and characterize metallic nanostructures.\cite{Sun10, Yan10} Due to the proceeding process of miniaturization the electrical and thermal properties of individual nanowires (NWs) are of major interest.\cite{Wu2004} The electrical and optical properties of silver nanowire (Ag NW) ensembles show a great potential for transparent conductors and touch screen applications.\cite{Groep12} For electrical interconnects, Ag NWs are promising because bulk silver possesses the highest electrical and thermal conductivity among metals.\cite{Wiley07} Certainly, a high thermal conductivity is of importance as a reduction of feature size or cross-section leads to higher current densities, higher resistances and higher power dissipation. Furthermore, the increase of the surface-to-volume ratio enhances the relative contribution of surface scattering in nanostructures. 

To date, individual single crystalline Ag NWs have been used as model systems to analyze electron transport and scattering mechanisms\cite{Huang09} as well as to study the effect of electromigration.\cite{Kaspers12} However, thermal transport measurements on Ag NWs were carried out only on ensembles of polycrystalline Ag NWs embedded in polycarbonate.\cite{Xu09} These measurements indicated a reduced thermal conductivity whereas grain boundary scattering dominated surface scattering due to grain sizes. Electrical and thermal conductivity measurements on other individual metallic NWs were previously reported.\cite{Volk08, Ou2008} However, the combination of thermoelectrical and structural characterization, as performed in this work, has been rarely applied for a single NW.\cite{Zhou05, Mavro09, Kojda2014} Furthermore, thermopower measurements have not been reported on Ag NWs, neither on ensembles nor on individual NWs. 

In this work, a comprehensive study of temperature-dependent thermoelectric and structural properties of high purity single crystalline Ag NWs is presented for the temperature range between \SI{1.4}{\kelvin} and room temperature (RT). In particular, we investigate the temperature dependence of the Lorenz number $L(T)$  and find $L(T)$ to be determined by the material's purity and to be independent of surface scattering. We observe the transition from diffusive transport to quasi ballistic one dimensional transport in a single metallic NW. 

\section{Experimental Details}

\subsection{Synthesis of Ag NWs}

Synthesis of Ag NWs has been carried out by various procedures, such as template based\cite{Huang00, Riveros06} and soft solution methods.\cite{Sun02, Zhou06}
Here, the Ag NWs were synthesized by reducing silver nitrate (\ce{AgNO3}) with ethylene glycol (EG) in presence of copper dichloride dihydrate (\ce{CuCl2 + 2H2O}) and polyvinylpyrrolidone (PVP). The EG is used as solvent and reducing agent whereas the PVP is used as capping agent.\cite{Zhang11} In the beginning, \SI{10}{\milli \liter} of EG (Sigma-Aldrich, anhydrous \SI{99.8}{\percent}) was heated in an oil bath at \SI{151.5}{\celsius} for \SI{1}{\hour} under continuous magnetic stirring at \SI{260}{\rpm}. Then, \SI{80}{\micro \liter} of a \SI{4}{\milli \molar} \ce{CuCl2 +2H2O} (Sigma-Aldrich, \SI{99.999}{\percent})/EG solution was added to the preheated EG and stirred for another \SI{15}{\minute}. Next, \SI{3}{\milli \liter} of \SI{0.282}{\milli \molar} PVP (Sigma-Aldrich, ${M_{\mathrm{w}} = \SI{1.3E6}{\gram \per \mol}}$) dissolved EG solution is added to the heated EG. Subsequently, \SI{3}{\milli \liter} of prepared \SI{0.094}{\milli \molar} \ce{AgNO3} (Sigma-Aldrich, \SI{99.9999}{\percent})/EG solution was injected by syringe pump with an injection speed of \SI{30}{\milli \liter \per \hour}. Before the injection, the \ce{AgNO3}/EG solution was sonicated for \SI{7}{\minute} as previously described.\cite{Lee12} The color of the resulting mixture changed from colorless to yellow ivory over dark grey to bright opaque gray. Upon NW formation the reaction was quenched by removing the flask from the oil bath. Thereafter, the solution containing suspended NWs was centrifuged and diluted by deionized water for several times.

\subsection{Functionality of the TNCP}

\begin{figure}[htbp]
  \includegraphics[width=0.95\columnwidth]{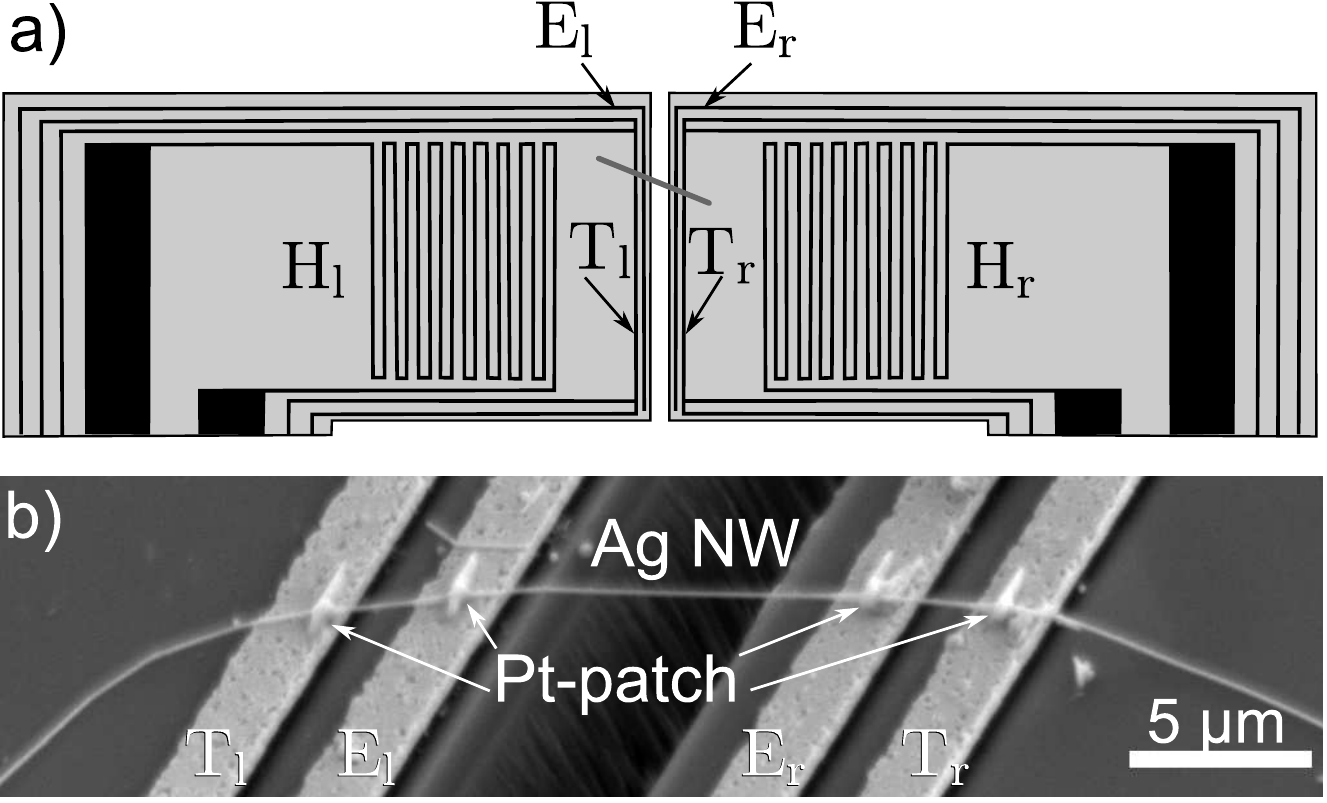}
  \caption{\label{fig:TNCP} (a) Sketch of the \textbf{T}hermoelectric \textbf{N}anowire \textbf{C}haracterization \textbf{P}latform (TNCP) cantilever tips (black -- Pt lines, bright gray -- \ce{SiO2}, dark gray -- NW, white -- vacuum). (b) SEM image of a tilted TNCP with an EBID contacted Ag NW (Ag NW3).}
\end{figure}

The thermoelectric nanowire characterization platform (TNCP) was produced by silicon micromachining.\cite{Wang13} This platform consists of two symmetric freestanding cantilevers with a minimum gap distance of \SI{4}{\micro \meter}. The electrical components such as meander-shaped micro heaters (H$_\mathrm{l}$, H$_\mathrm{r}$), resistance thermometers in four-point arrangement (T$_\mathrm{l}$, T$_\mathrm{r}$) and additional electrodes (E$_\mathrm{l}$, E$_\mathrm{r}$) are depicted in Fig.~\ref{fig:TNCP}a. 

These components were created on top of an isolating \SI{500}{\nano \meter} thick \ce{SiO2} layer by radio frequency sputtering of titanium (\SI{10}{\nano\meter}) and platinum (\SI{200}{\nano\meter}). The electrodes on the cantilever tips allow four-terminal sensing for the thermoelectric transport measurements, whereas current is injected via the thermometer electrodes T$_\mathrm{l}$ and T$_\mathrm{r}$ and the voltage drop at the NW is probed at the inner electrodes E$_\mathrm{l}$ and E$_\mathrm{r}$. The thermovoltage is measured between E$_\mathrm{l}$ and E$_\mathrm{r}$, whereas the temperature difference between the cantilevers is determined by a four-terminal resistance measurement of the calibrated lines T$_\mathrm{l}$ and T$_\mathrm{r}$. The TNCP fits the geometric requirements of conventional TEM sample holders (i.e. the maximum lateral size of less than \SI{3}{\milli \meter} and maximum height of about \SI{170}{\micro \meter}) allowing the chemical and structural analysis of the same NW immediately after the thermoelectric transport measurements. 

\subsection{Nanowire Assembly --- Electrical and Thermal Contacts}

Water suspended Ag NWs were dropped on a grooved structure of photoresist prepared by means of laser lithography. Herefrom, individual NWs were lifted and placed between the TNCP cantilevers with a thin indium tip.\cite{Flohr11} In order to ensure that the Ag NW touches all the Pt contacts the NW was pushed towards the platinum lines by the indium tip. The electrical contacts were checked with Keithley SourceMeter 2401 (voltage mode) in two point configuration and a series resistance of \SI{1}{\giga \ohm} was applied in order to protect the NW. The ohmic contact between the NW and the Pt lines was created at an applied voltage of \SI{1}{\volt}, when pushing the NW with the indium tip until a current was measured. 

Even though electrical contacts can be established by the described method the resulting contact area of the NW to the TNCP substrate is small which can lead to unfavorably high thermal resistances. In order to reduce the thermal resistance between the NW and the TNCP, electron beam induced deposition (EBID) of platinum contacts was established. The Nova 600 NanoLab (FEI) chamber was fed by the metal organic precursor \ce{(CH3)3Pt(C_pCH3)} which was cracked by a \SI{10}{\kilo \volt} electron beam at about \SI{2.4}{\nano \ampere} specimen current resulting in a platinum-carbon combustion covering the wire within a defined pattern.\cite{Botman08} Fig.~\ref{fig:TNCP}b shows a SEM image taken after the deposition, where \ce{Pt} patches smoothly cover the Ag NW and define the contact area. In the following we distinguish NWs without EBID contacts as \lq{}as-assembled NWs\rq{} and those with EBID contacts as \lq{}EBID-contacted NWs\rq{}.

\subsection{Measurement Procedure}

The structure of the NWs was investigated by transmission electron microscopy (TEM) using a JEOL JEM2200FS microscope operating at \SI{200}{\kilo \volt}. In particular, water suspended Ag NWs were dropped onto a copper (Cu) grid and left in ambient lab conditions for about one week before starting the TEM analysis. This experiment allowed analyzing structural changes in the NWs due to their exposure to air. Additionally to this experiment, individual Ag NWs positioned and processed on TNCPs were characterized by TEM after the measurement of the thermoelectric properties. Conventional TEM, high-angle annular dark-field scanning transmission electron microscopy (HAADF STEM) as well as energy dispersive X-ray (EDX) spectroscopy were applied for the structural analysis.

Thermoelectric transport measurements were performed in a flow cryostat in helium (\ce{He}) atmosphere at ambient pressure (Ag NW1 and Ag NW2) and in vacuum (Ag NW3, Ag NW4; $p < \SI{5E-6}{\milli \bar}$) at bath temperature~\TB{}. \TB{} was measured by Cernox\textsuperscript{\texttrademark} sensor placed nearby the TNCP and was varied in the range between \SI{1.4}{\kelvin} and RT. The measurements were carried out for an absolute temperature stability of $\pm$\SI{50}{\milli \kelvin}. In order to protect the wire from external voltage peaks two symmetrical low-pass filters (\SI{1.6}{\kilo \ohm} + \SI{1}{\micro \farad} + \SI{1.6}{\kilo \ohm}) were connected to the current inputs T$_\mathrm{l}$ and T$_\mathrm{r}$. For electrical conductivity current-sweeps up to a maximum current of $I_{\mathrm{max}} = \SI{0.5}{\milli \ampere}$ were performed in four-terminal sensing by Keithley 2401. To cancel the effect of the low pass filters, the voltage measurement was taken \SI{1}{\second} after the current was set. For Seebeck coefficient measurements the voltage drop between E$_\mathrm{r}$ and E$_\mathrm{l}$ as well as the resistances of the thermometers T$_\mathrm{l}$ and T$_\mathrm{r}$ were measured as function of the applied heater current and the bath temperature \TB{}. AC measurements for thermal conductivity were taken by lock-in amplifier SR830 without additional low pass filters, but with a series resistance of \SI{10}{\kilo \ohm} for current limitation. 

\section{Results}

\subsection{TEM-Analysis of Ag NWs on a Cu Grid}
\label{sec:TEM_on_grid}

\begin{figure}[htbp]
  \includegraphics[width=\columnwidth]{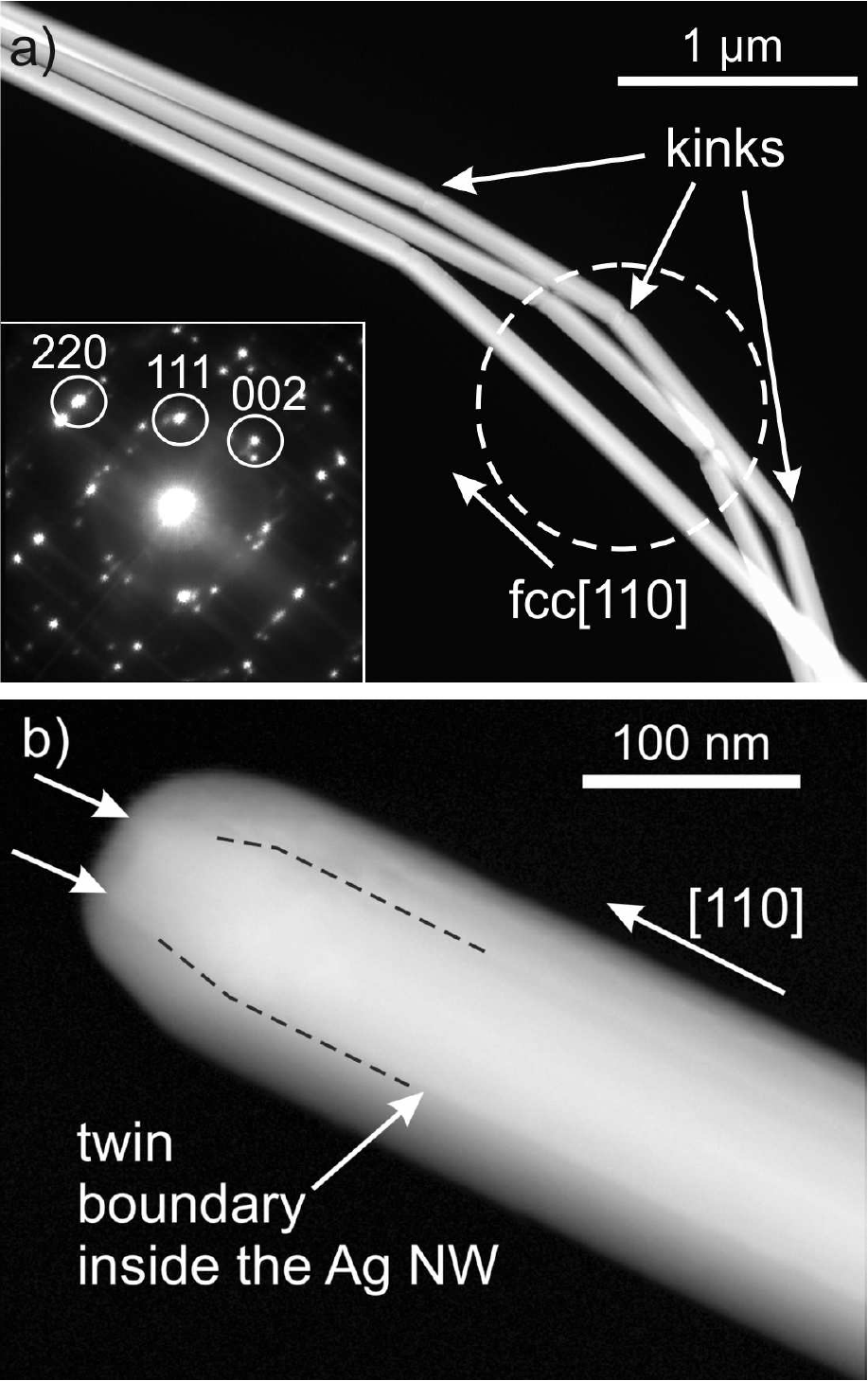}
  \caption{\label{fig:TEM1}(a) HAADF STEM micrograph showing Ag NWs with the rare case of kinks. The inset shows electron diffraction pattern obtained from a bunch of NWs (see the region marked by a dashed line in the STEM image); (b) HAADF STEM image showing a residual diffraction contrast due to the presence of a twin boundary  along the growth direction of an Ag NW.}
\end{figure}

TEM analysis of Ag NWs dropped onto a Cu grid has shown that the PVP-mediated polyol process leads to formation of NWs with diameters ranging from \num{88} to \SI{135}{\nano \meter}. Straight NWs with lengths exceeding \SI{35}{\micro \meter} were observed. Similarly to the previously published studies\cite{Liu06} electron diffraction analysis confirms the presence of fcc structure and the NW growth direction corresponds to Ag[110], as shown in the inset of Fig.~\ref{fig:TEM1}a. HAADF STEM image in Fig.~\ref{fig:TEM1}a shows the presence of kinks along the NWs. Kink defects may act as scattering centers during the electron and phonon transport through the NWs. The distance between the kinks is irregular and varies from hundreds of nanometers to a few micrometers. At the kink positions the structure is rotated around the [110] growth direction and the [110] growth axis is tilted with respect to the previous NW segment. Thus, the NW growth axis remains the same for the different NW segments, i.e. Ag[110], but the spatial geometry of the NWs changes. The misalignment of the Ag[110] NW growth axis between the attached NW segments suggests the presence of grain boundaries at the kink positions. These crystal defects might be randomly formed during the NW growth process.

Furthermore, the NWs consist of twin regions with twin boundaries arranged along the NW growth axis. Figure~\ref{fig:TEM1}b shows a HAADF STEM image with a residual diffraction contrast appearing at the twin plane which intersects the Ag NW along its growth direction. A similar finding has been previously reported.\cite{Sun03} The presence of twins in the Ag NWs can be understood considering the structure of Ag nanoparticles used as seeds for the NW growth. These nanoparticles typically contain twin boundaries. Sun \textit{et al.} demonstrated the 5-fold twinned structure of Ag nanoparticles formed prior to the formation of Ag NWs of pentagonal shape.\cite{Sun03} During the NW formation these Ag-seeds with a multi-twinned structure grow preferentially along one of the $<$110$>$ directions due to the preferential Ag atom attachment at twin boundaries. In contrast, lateral NW growth is nearly suppressed due to the passivation of the side facets of the Ag seeds by the PVP molecules. As result, the multi-twinned structure of the Ag seeds is preserved in the NWs.

\begin{figure}[htbp]
  \includegraphics[width=\columnwidth]{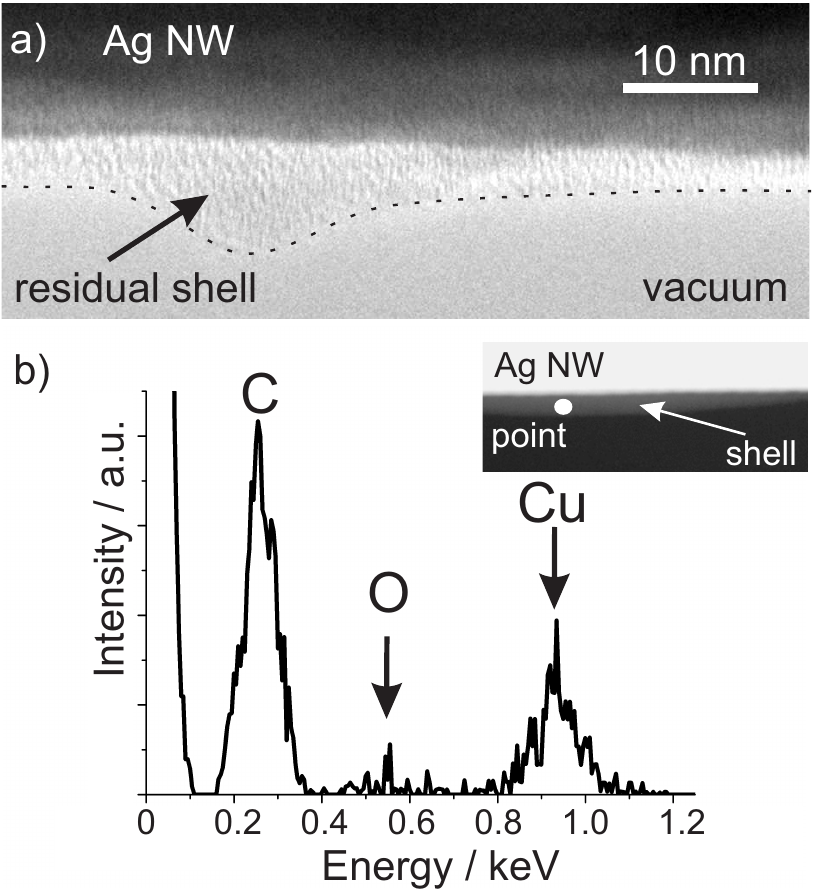}
  \caption{\label{fig:TEM2a} (a) TEM micrograph showing an amorphous shell on the NW surface. The interface between the shell and the vacuum is indicated by a dotted line. (b) EDX spectrum indicating the presence of carbon in the surface layer. A HAADF STEM image in the inset shows the NW surface with indicated beam position during the measurement.}
\end{figure}

Two different surface features were observed at the Ag NWs. First, the NWs are covered by an amorphous film with a thickness varying from \num{2} to \SI{9}{\nano \meter} (Fig.~\ref{fig:TEM2a}a). According to the EDX analysis this film contains a high amount of carbon (C) (Fig.~\ref{fig:TEM2a}b). It cannot be excluded that a certain amount of oxygen (O) is also present in this surface layer, although the oxygen peak intensity in the EDX spectrum is rather low compared to that of the carbon peak. The Cu--L peak visible in the EDX spectrum in Fig.~\ref{fig:TEM2a}b appears as spurious signal due to the presence of the Cu grid. The formation of the C-containing surface film can be attributed to the PVP capping used during the NW synthesis process as has been shown elsewhere.\cite{Lee12} Hence, the NWs consist of a core-shell structure with an Ag NW as a core and most probably PVP as a shell.

\begin{figure}[htbp]
  \includegraphics[width=\columnwidth]{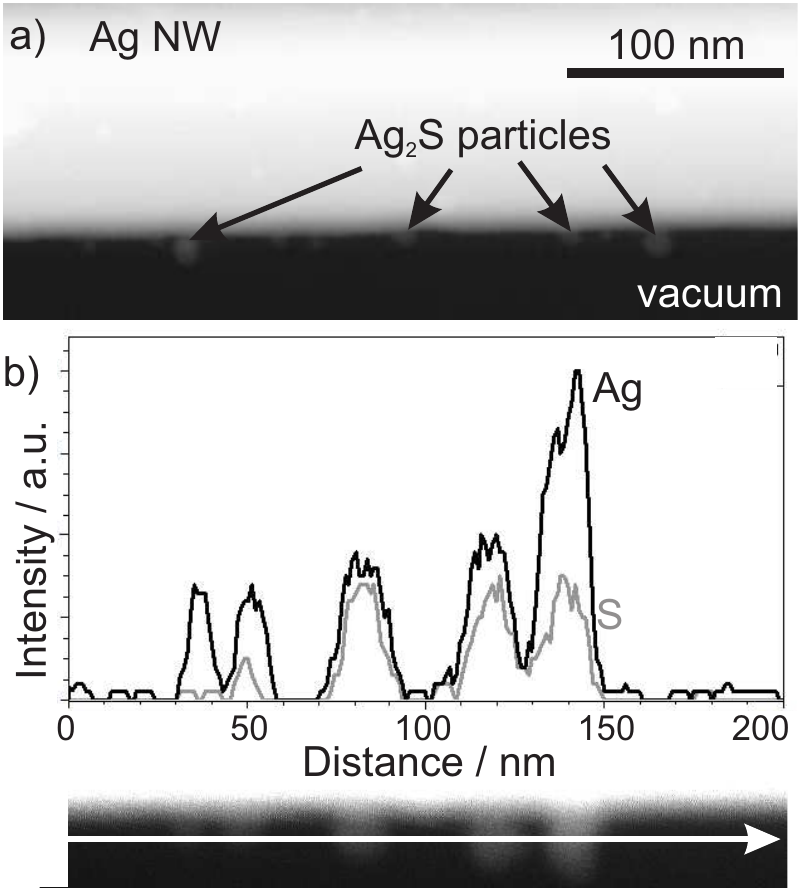}
  \caption{\label{fig:TEM2b} (a) HAADF STEM image of the NW surface showing the presence of round-shaped particles. (b) EDX line scan across the surface (see the arrow in the HAADF STEM inset) proving the presence of Ag and S in the surface particles.}
\end{figure}

The second observed surface feature is shown in Fig.~\ref{fig:TEM2b}a. The HAADF STEM micrograph reveals the formation of crystallites with a round shape on the NW surface. EDX analysis unambiguously proves the presence of sulfur and silver in these regions (see the line scan in Fig.~\ref{fig:TEM2b}b). Consequently, the round-shaped surface particles represent \ce{Ag2S}, which can be formed on the Ag surface due to its exposure to the air.\cite{Cao09} The Ag sulfide is known to have low electrical conductivity.\cite{Hebb52} Indeed, we observed strong charging effects during TEM observation of Ag NWs covered with \ce{Ag2S} particles, and the charging increased with a rising number of \ce{Ag2S} nanoparticles. The formation of Ag sulfide can be attributed to the Ag sulfidation which is typical for bulk silver exposed to the air. For this reason the individual positioned, processed and electrically characterized Ag NWs on TNCPs were kept in a \ce{N2} atmosphere before being analyzed by TEM.

\subsection{TEM-Analysis of Ag NWs on the TNCP}

\begin{figure*}
  \includegraphics[width=\textwidth]{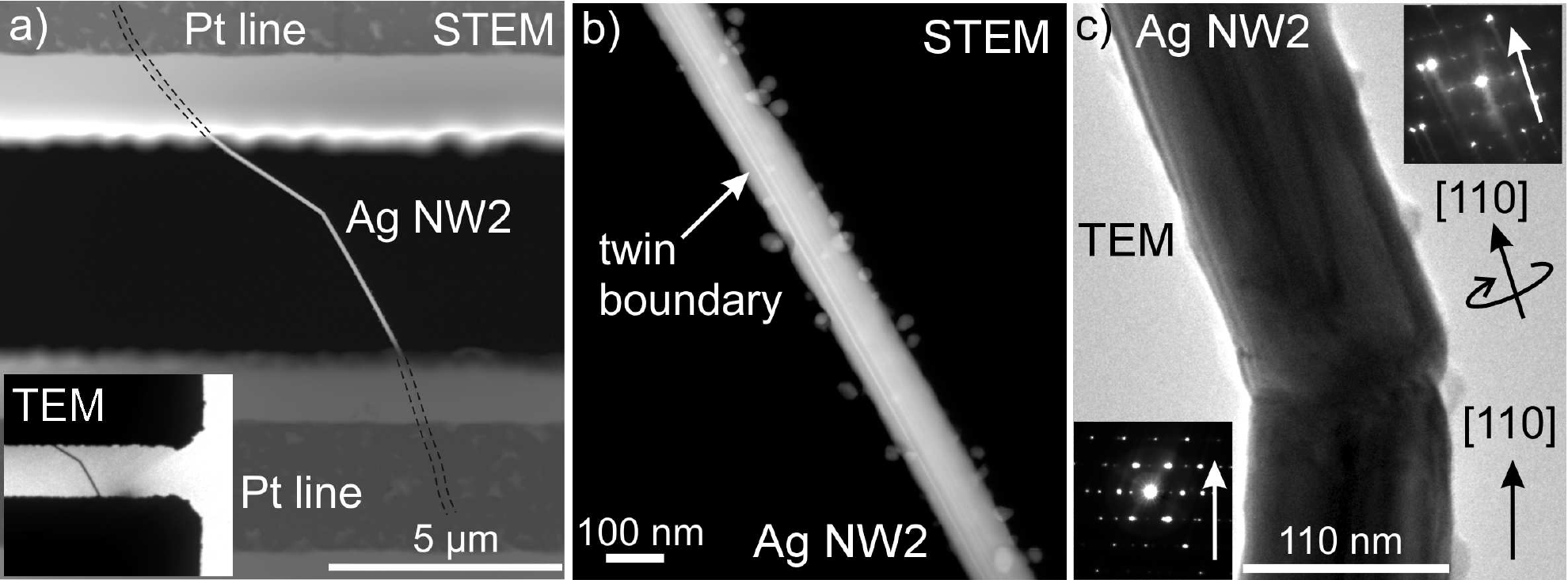}%
  \caption{\label{fig:TEM3} (a) STEM micrograph of a topview of Ag NW2 on the TNCP. The inset shows the corresponding TEM image. (b) HAADF STEM image of a segment of Ag NW2. (c) TEM image of a kink at the NW2. The insets show selected area electron diffraction (SAED) patterns of the corresponding NW segments.}
\end{figure*}

Next to structural and chemical characterization of Ag NWs on Cu grids, we took great care to analyze Ag NWs by TEM methods which were thermoelectrically investigated before. In particular, we present the results of Ag NWs on top of the TNCP. After measuring the thermoelectrical properties of the NWs on the TNCPs, they were loaded into a single-tilt TEM specimen holder and analyzed by TEM. Figure~\ref{fig:TEM3} shows an exemplary Ag NW (Ag NW2) on a TNCP. Conventional TEM allowed analysis of the NW between the TNCP gap (see inset in Fig.~\ref{fig:TEM3}a). STEM with a HAADF detector allowed the investigation of a topview of the NW on the TNCP (Fig.~\ref{fig:TEM3}a). The TNCP cantilevers were intransparent to electrons due to a thickness of about \SI{20}{\micro \meter}. However, the presence of the gap in the TNCP allowed electrons scattered from the cantilever to pass through the gap and to reach the HAADF detector. As a result, it is possible to obtain a certain topologic contrast from TNCPs despite their high thickness. Since the contrast of the NW on the intransparent TNCPs is rather low, the NW position on the TNCP was marked by black dashed lines in Fig.~\ref{fig:TEM3}a. 

Figure~\ref{fig:TEM3}b shows an exemplary HAADF STEM micrograph of a segment of Ag NW2 in the TNCP gap. Here, round features on the NW surface contain Ag. Therefore, these particles have to be distinguished from those originating by the Ag sulfidation process as described in section~\ref{sec:TEM_on_grid}. Small hutches under the silver particles result in additional roughing of the surface. This may result from electromigration during the thermoelectrical measurements at higher current densities. However, this kind of Ag particles was not observed for Ag NW4, which revealed a smooth surface. 

As described in section~\ref{sec:TEM_on_grid} for Ag NWs on Cu grids, the thermoelectrically characterized Ag NWs on the TNCP contain twin boundaries. An example is shown in Fig.~\ref{fig:TEM3}b, where the twin boundary is indicated by the bright line along the NW growth direction. A single pronounced kink as visible in Fig.~\ref{fig:TEM3}a and Fig.~\ref{fig:TEM3}c was only observed for Ag NW2. The TEM micrograph in Fig.~\ref{fig:TEM3}c clearly shows that both NW segments intergrow at the kink position, which enables electron transport through the NW. This finding proves the above suggestion that the kinks appear during the NW growth process and not as a result of NW assembly or mechanical stress.

\subsection{Temperature-dependent Electrical Conductivity}
\label{sec:ElectricalConductivity}

\begin{figure}[htbp]
  \includegraphics[width=\columnwidth]{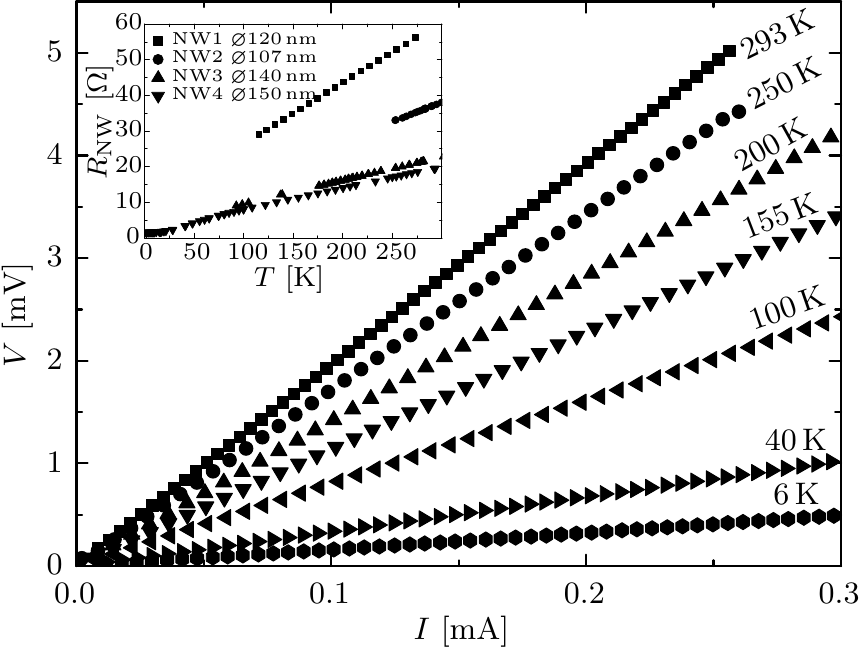}
  \caption{\label{fig:Resistance} Four point measured $I$-$V$ curves of an individual Ag NW (Ag NW4) at different bath temperatures. The inset shows the measured resistances for the Ag NWs with diameters ranging from \SI{107}{\nano\meter} to \SI{150}{\nano\meter}.}
\end{figure}

For the electrical characterization of the Ag NWs, DC and AC four-terminal sensing measurements were performed. The measured current-voltage ($I$-$V$) characteristics of the NWs show ohmic behavior as exemplarily depicted in Fig.~\ref{fig:Resistance}. The inset of Fig.~\ref{fig:Resistance} shows the NWs' resistances  exhibiting metallic behavior.  

The resistance of NW4 was measured in the temperature range between \SI{1.4}{\kelvin} and RT. Therefrom results the residual resistance ratio $RRR = \frac{R_\mathrm{NW}(\RT{})}{R_\mathrm{NW}(T=\SI{1.4}{\kelvin})} \approx 12$. With respect to pure bulk samples the $RRR$s in Ag NWs show strongly reduced values. This is due to the fact, that the bulk electron mean free path (EMFP) at low $T$ is determined by impurity scattering, whereas the EMFP in NWs is dominated by size dependent electron surface scattering. Therefore, the $RRR$ in NWs is no absolute measure of crystal quality as it is for the bulk. 

\begin{figure}[htbp]
  \includegraphics[width=\columnwidth]{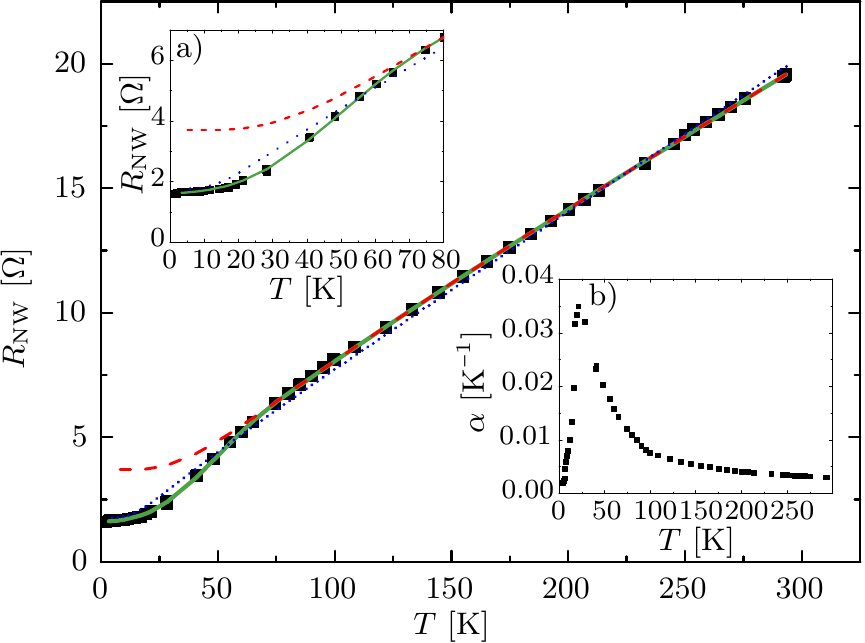}
  \caption{\label{fig:Bloch} Measured resistance of Ag NW4 fitted by Eq. \eqref{eq:Bloch} and \eqref{eq:BlochX}. Best fit to the measurement data is given by Eq. \eqref{eq:BlochX} with $\Theta_\mathrm{D} = \SI{215}{\kelvin}$, $R_0 \rightarrow \SI{1.62}{\ohm}$, $T_\mathrm{SS} \rightarrow \SI{49.4}{\kelvin}$, $M \rightarrow \SI{2.1}{\ohm}$ and $\Reph{} \rightarrow \SI{48}{\ohm}$ shown by the green solid line. The dashed red line shows the fit of Eq. \eqref{eq:Bloch} for fixed $\Theta_{\mathrm{D}}=\SI{215}{\kelvin}$ with $\Reph{} = \SI{48}{\ohm}$ and $R_0 = \SI{3.7}{\ohm}$ . The dotted blue line shows the fit for Eq. \eqref{eq:Bloch} with $\Theta_\mathrm{D} = \SI{86}{\kelvin}$ and $R_0 \rightarrow \SI{1.62}{\ohm}$. Inset (a) shows the measured resistances and fits for lower temperatures. Inset (b) shows the TCR $\alpha$ of NW4 as function of temperature.}
\end{figure}

Generally, the temperature dependence of the resistance $R(T)$ of a metal can be described by the Bloch-Gr\"uneisen formula given by Eq. \eqref{eq:Bloch}.\cite{Ziman1996} 
\begin{equation}
  R_\mathrm{BG}(T) = R_0 + \Reph{} \left( \frac{T}{\Theta_\mathrm{D}}\right)^5 \int_0^{\frac{\Theta{}_\mathrm{D}}{T}}{\frac{x^5 e^x}{(e^x-1)^2}} \mathrm{d}x 
\label{eq:Bloch}
\end{equation}
Herein, $R_0$, $\Reph{}$ and $\Theta{}_\mathrm{D}$ describe the residual resistance, a scalar determining the electron-phonon coupling and the Debye temperature, respectively. Initially, the measured resistances of NW4 were fitted with respect to Eq. \eqref{eq:Bloch} as shown in Fig.~\ref{fig:Bloch}. For the fixed Debye temperature of silver\cite{Smith95} $\Theta_\mathrm{D, Ag} = \SI{215}{\kelvin}$ the data points above \SI{100}{\kelvin} are well described whereas the data at low temperatures deviate from the theory (dashed line). In Ref. \onlinecite{Bid2006} this problem was solved by fitting the measured resistances of Ag NW ensembles with lower Debye temperatures. Varying the parameter $\Theta_\mathrm{D}$, the fit for the function $R_\mathrm{BG}(T)$ yields $\Theta_\mathrm{D} \rightarrow \SI{86}{\kelvin}$ (dotted line in Fig.~\ref{fig:Bloch}) also indicating a reduction in the Debye temperature. But the relative error of this fit constitutes up to \SI{20}{\percent} at about $T=\SI{30}{\kelvin}$, so that it failed to describe $R(T)$ at low temperatures (compare dotted line in the inset (a) of  Fig.~\ref{fig:Bloch}). In particular, a strong reduction of $\Theta_\mathrm{D}$ lacks any physical explanation.
 
For a more accurate treatment one has to consider that the transport regime changes from diffusive transport to quasi ballistic transport with decreasing $T$. As discussed later in detail, the EMFP exceeds the diameter $d$ in the regime of quasi ballistic transport, so that a high fraction of electrons reaches the surface. Hence, electron scattering is dominated by surface scattering. Furthermore, small angle scattering dominates the electron-phonon scattering events.\cite{Tritt04} In the following we deduce a semi-empirical expression for the temperature-dependent influence of surface scattering on the measured resistance $R_\mathrm{NW}(T)$ based on ideas found in Ref. \onlinecite{Tritt04}.  

The surface scattering gives rise to an additive resistance component $R_\mathrm{SS} \propto \Lambdael{,SS}^{-1}$, where \Lambdael{,SS} is the electron mean free path for surface scattering. Electrons, that are scattered by the angle $\gamma$ contribute to the resistance only when they are backscattered. Thus, the probability of backscattering determines the NW's resistance. Therefore, the differential cross-section for backscattering $\phi_\mathrm{diff}$, that depends on the NW's geometry, is responsible for surface scattering. Due to $\Lambdath{,SS}^{-1} \propto \phi_\mathrm{diff} \Omega$, $R_\mathrm{SS}$ scales with the solid angle $\Omega$ that is defined by the cone angle $2 \gamma$. The solid angle $\Omega$ relates to the cone angle $2 \gamma$ with $\Omega = 2\pi (1-\cos(\gamma))$.\cite{Tritt04} With the latter relation and $\gamma \propto T$ one finds the semi-empirical expression $R_\mathrm{SS} \propto \Omega \propto \sin^2(T/T_\mathrm{SS})$ for the temperature dependence of surface scattering. Here, $T_\mathrm{SS}$ defines the threshold temperature for which surface scattering becomes dominant.

In order to describe the complete $R_\mathrm{NW}(T)$ characteristic, the additional scattering term $R_\mathrm{SS} \propto \sin^2\left(\frac{T}{T_\mathrm{SS}}\right)$ describing the influence of surface scattering was added for the low temperature regime as described by Eq. \eqref{eq:BlochX}
\begin{align}
  R_\mathrm{NW}(T) &= R_\mathrm{BG}(T) + R_\mathrm{SS}(T) \notag \\ 
	R_\mathrm{SS}(T) &= \begin{cases} M &\mbox{if }  \frac{T}{T_\mathrm{SS}} > \frac{\pi}{2} \\
                                    M \sin^2\left(\frac{T}{T_\mathrm{SS}}\right) & \mbox{else} \end{cases}, 	
	\label{eq:BlochX}
\end{align}
where $M$ represents the strength of scattering.  

As shown in Fig.~\ref{fig:Bloch}, the resulting fit compares well with the data. We want to point out, that Eq.~\ref{eq:BlochX} describes our data for the bulk value $\Theta_\mathrm{D,Ag}$, so that a consideration of a reduction in the Debye temperature $\Theta_\mathrm{D}$ is not necessary. Due to the low appearance of grain boundaries, grain boundary scattering will not change the resistance significantly and can be neglected in our consideration. The same applies to the twin boundaries as they are known to be weak scatterers.\cite{Huang09} However, the twins' alignment parallel to the NW's surface and parallel to the direction of transport possibly act as additional confinement. 

The temperature coefficients of resistance (TCR) $\alpha = \frac{1}{\RW{}} \frac{\partial \RW{}}{\partial \TB{}}$ of the NWs as well as their temperature dependence were found to compare with each other. In the inset b) of Fig.~\ref{fig:Bloch} the TCR of NW4 is shown as a function of $T$. Below \SI{100}{\kelvin} $\alpha(T)$ shows a strong increase with $T$ and a maximum at \SI{21}{\kelvin}. Comparing the $\alpha(\mathrm{RT})$ with the silver bulk value of $\alpha_\mathrm{B}(\RT{})=\SI{3.79E-3}{\per \kelvin}$ the measured values for Ag NWs are reduced by about \SI{25}{\percent}.\cite{Seth70} A similar reduction in $\alpha$ was observed in electrodeposited Ag NWs.\cite{Bid2006}

The electrical conductivity $\sigma(T)$ of the Ag NWs was calculated by ${\sigma(T) = 4 \lw{}/(\RW{}(T) \pi d^2)}$ by assuming cylindrical wire cross-sections. The values $\lw{}$ and $d$ represent the NWs' lengths and their diameters respectively. The NWs' geometries were determined by SEM imaging (Ag NW3) and by STEM imaging (Ag NW1, Ag NW2, Ag NW4). The resulting $\sigma (\mathrm{RT})$, the NWs' geometries and $\alpha(\mathrm{RT})$ are given in Table~\ref{tab:conductivity}. The calculated uncertainties in $\sigma$ result from diameter variations which are due to image resolution and due to a surface roughness. The high uncertainty of the diameter of Ag NW1 results from the STEM uncertainties that occur in this case.  

\begin{table}[htbp]
  \caption{\label{tab:conductivity} Electrical conductivity $\sigma$, TCR $\alpha$ and the NW geometry of different Ag NWs at RT.}
  \begin{ruledtabular}
	  \begin{tabular}{lllll}
		  $\sigma [\SI{E7}{\siemens \per \meter}]$ & $\alpha [\SI{E-3}{\per \kelvin}]$ &  $d [\si{\nano \meter}]$ & $\lw{} [\si{\micro \meter}]$  & Sample No. \\ \hline
      \num{3.1 \pm 1.1} & \num{2.86 \pm 0.02}  & \num{120 \pm 20} & \num{15 \pm 1} & Ag NW1   \\
      \num{3.5 \pm 0.6} & \num{2.88 \pm 0.02}  & \num{107 \pm 5}  & \num{11 \pm 1} & Ag NW2   \\
      \num{3.8 \pm 0.8} & \num{2.94 \pm 0.02}  & \num{140 \pm 10} & \num{13 \pm 1} & Ag NW3   \\
			\num{4.0 \pm 0.2} &  \num{2.95 \pm 0.02} & \num{150 \pm  3} & \num{14 \pm 1} & Ag NW4   \\	
    \end{tabular}
  \end{ruledtabular}
\end{table}

In comparison to the electrical conductivities of polycrystalline bulk silver (purity: \SI{99.9999}{\percent},  ${RRR=1800}$) of $\sigma_\mathrm{B}=\SI{6.16E7}{\siemens \per \meter}$ the measured conductivities of Ag NWs are reduced by \SI{35}{\percent} to \SI{50}{\percent}.\cite{Seth70} Reduced electrical conductivities in Ag NWs of \SI{40}{\nano\meter} diameter and \SI{50}{\nano\meter} diameter prepared by the polyol method were previously reported.\cite{Sun02b, Peng11} Within the measurement uncertainty, the measured $\sigma(\RT{})$ compare with cross-section dependent calculations on thin Ag NWs.\cite{Huang09} 

\begin{figure}[htbp]
  \includegraphics[width=\columnwidth]{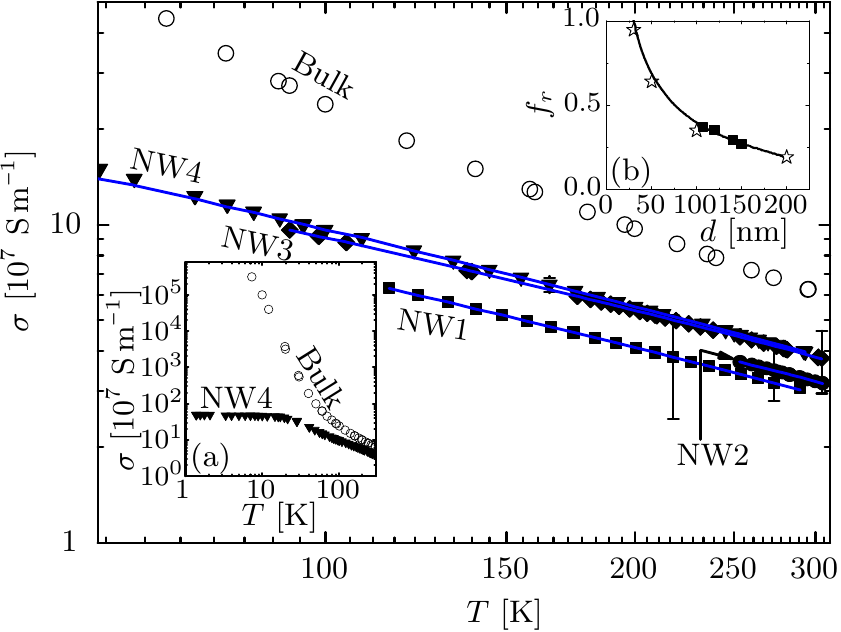}
  \caption{\label{fig:sigma} Electrical conductivity $\sigma_\mathrm{NW}(T)$ of individual Ag NWs (NW1: $\filledmedsquare$; NW2: {\Large $\bullet$}; NW3: {\Large $\filleddiamond$}; NW4: {\Large $\filledtriangledown$}) as function of the temperature \TB{}. Fitted curves of Eq. \eqref{eq:Matthiessen} are depicted as solid line. Reference values of Ag bulk are denoted by ({\Large $\circ$}).\cite{Seth70, White91} (a) Electrical conductivity of NW4 compared to the bulk reference ({\Large $\circ$}) in the measured temperature region. (b) Fraction $f_r$ of NWs of this study ($\filledmedsquare$) compared to literature values on NW ensembles ($\largestar$).\cite{Bid2006}}
\end{figure}

The temperature dependence $\sigma(T)$ of the Ag NWs is shown in Fig.~\ref{fig:sigma}. Generally, the EMFP $\Lambdael{,NW}(\TB)$ of the NWs can drop with respect to the bulk due to enhanced surface scattering. A reduced scattering length can be considered in terms of Matthiessen rule by Eq. \eqref{eq:Matthiessen}
\begin{equation}
  \Lambdael{,NW}(\TB)^{-1} = \Lambdael{,B}(\TB)^{-1} + \Lambdael{,SS}^{-1},
  \label{eq:Matthiessen}
\end{equation}
with $\Lambdael{,B}(\TB)$ the temperature-dependent bulk EMFP and $\Lambdael{,SS}$ the temperature independent scattering length due to the surface scattering. Since $\sigma \propto \Lambda$, relative changes in electrical conductivity correlate with relative changes in EMFP as described for the bulk and the NW by Eq. \eqref{eq:LambdasI} and Eq. \eqref{eq:LambdasII}, respectively. 
\begin{align}
  \Lambdael{,B}(\TB) &= \frac{\sigma_\mathrm{B}(\TB)}{\sigma_\mathrm{B}(\RT{})} \Lambdael{,B}(\RT{})      \label{eq:LambdasI} \\ 
	\Lambdael{,NW}(\TB) &= \frac{\sigma_\mathrm{NW}(\TB)}{\sigma_\mathrm{NW}(\RT{})} \Lambdael{,NW}(\RT{})  \label{eq:LambdasII}
\end{align}

By substituting Eq. \eqref{eq:LambdasI} and Eq. \eqref{eq:LambdasII} into Eq. \eqref{eq:Matthiessen} one can fit the relative changes of the NWs' electrical conductivities to determine the free parameters $\Lambdael{,NW}(\RT{})$ and $\Lambdael{,SS}$. For this purpose, the bulk reference values\cite{Seth70, White91} for $\sigma_\mathrm{B}(\TB)$ and the RT EMFP\cite{Stojanovic10} of $\Lambdael{,B}(\RT{}) = \SI{53}{\nano\meter}$ were taken. The resulting fit parameters are given in Tab.~\ref{tab:fit}, whereas the corresponding fit functions are shown in Fig.~\ref{fig:sigma}. These results are valid in the diffusive transport regime ($T\gtrsim T_\mathrm{SS}$), at which the electron-phonon scattering and  surface scattering contribute. Here, \Lambdael{,SS} is constant with respect to $T$. 

\begin{table}[htbp]
  \caption{\label{tab:fit} Determined fit parameters of Eq. \eqref{eq:Matthiessen} and resulting values of $f_r$.}
  \begin{ruledtabular}
	  \begin{tabular}{lllll}
      $f_r$ &  $d [\si{\nano \meter}]$   &   $\Lambdael{,NW}(\RT{})$ & $\Lambdael{,SS}$ & Sample No. \\ \hline
      \num{0.35}  &  \num{120 \pm 20}          & \SI{42}{\nano\meter}      & \SI{186}{\nano\meter}  & Ag NW1 \\
      \num{0.37}  &  \num{107 \pm  5}          & \SI{40}{\nano\meter}      & \SI{161}{\nano\meter}  & Ag NW2\\
			\num{0.29}  &  \num{140 \pm 10}          & \SI{41}{\nano\meter}      & \SI{182}{\nano\meter}  & Ag NW3\\
			\num{0.27}  &  \num{150 \pm  3}          & \SI{41}{\nano\meter}      & \SI{210}{\nano\meter}  & Ag NW4\\
    \end{tabular}
  \end{ruledtabular}
\end{table}

We observe reduced $\Lambdael{,NW}(\RT{})$ with respect to $\Lambdael{,B}(\RT{})$. The fractions $f_r=\Lambdael{,NW}(\RT{})/d$ are in agreement with the Ag NW ensemble measurements\cite{Bid2006} as shown in the inset (a) of Fig.~\ref{fig:sigma}. Hence, we confirm that the NWs' EMFPs behave as theoretically considered for \RT{}.\cite{Dingle1950} The resulting scattering lengths $\Lambdael{,SS}$ are in the order of the NWs' diameters confirming the reduced conductivities to be due to the NW size. 

\subsection{Temperature-dependent Seebeck Coefficient}
\begin{figure}[htbp]
  \includegraphics[width=\columnwidth]{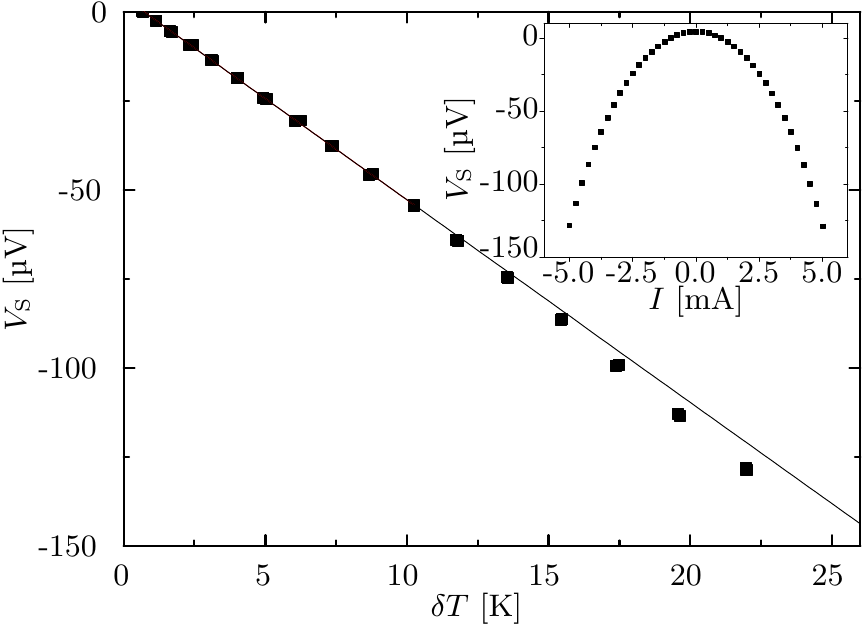}
  \caption{\label{fig:Seebeck} Thermovoltage as function of the temperature difference $\delta T$ between the cantilevers and (inset) thermovoltage as function of the heater current at RT.}
\end{figure}

The thermovoltage \VS{} of as-assembled Ag NWs (NW on top of Pt lines and without EBID contacts) was measured in helium atmosphere as function of the heater current $I$ in the temperature range between \SI{150}{\kelvin} and RT. The inset of Fig.~\ref{fig:Seebeck} shows the measured \VS{} as function of $I$ at \RT{}. The parabolic behavior of \VS{} as well as the observed change of sign by changing the heater side confirm the measurement of a thermoelectric voltage. The temperature difference $\delta T$ between both TNCP cantilevers was determined from the measured resistance change of calibrated thermometer lines T$_{\mathrm{l}}$ and T$_{\mathrm{r}}$. Hence, \VS{} can be analyzed as function of $\delta T$, as shown in Fig.~\ref{fig:Seebeck}. 

From the slope of $V_\mathrm{S}(\delta T)$ the thermopower of Ag with respect to Pt, $S_\mathrm{Ag, Pt}$, can be determined. In accordance to the measurement configuration (reference voltage on the cold side, Ag NW placed between Pt electrodes) the Seebeck coefficient is given by $S_{\mathrm{Ag, Pt}} = S_{\mathrm{Ag}} - S_{\mathrm{Pt}} = -\mathrm{d}\VS{}/\mathrm{d}\delta T$, with $S_{\mathrm{Ag}}$ and $S_{\mathrm{Pt}}$ the absolute Seebeck coefficients of \ce{Ag} and \ce{Pt}. Hence, the RT Seebeck coefficient yields $S_{\mathrm{Ag, Pt}} = \SI{5.9\pm0.7}{\micro \volt \per \kelvin}$. Due to the temperature dependency of $S_{\mathrm{Ag, Pt}}(T)$ the slope $-\mathrm{d}\VS{}/\mathrm{d}\delta T$ has to be calculated for small $\delta T$. For data analysis we chose $\delta T < \SI{10}{\kelvin}$ in order to balance the latter condition with a significant signal in $V_\mathrm{S}$.

\begin{figure}[htbp]
  \includegraphics[width=\columnwidth]{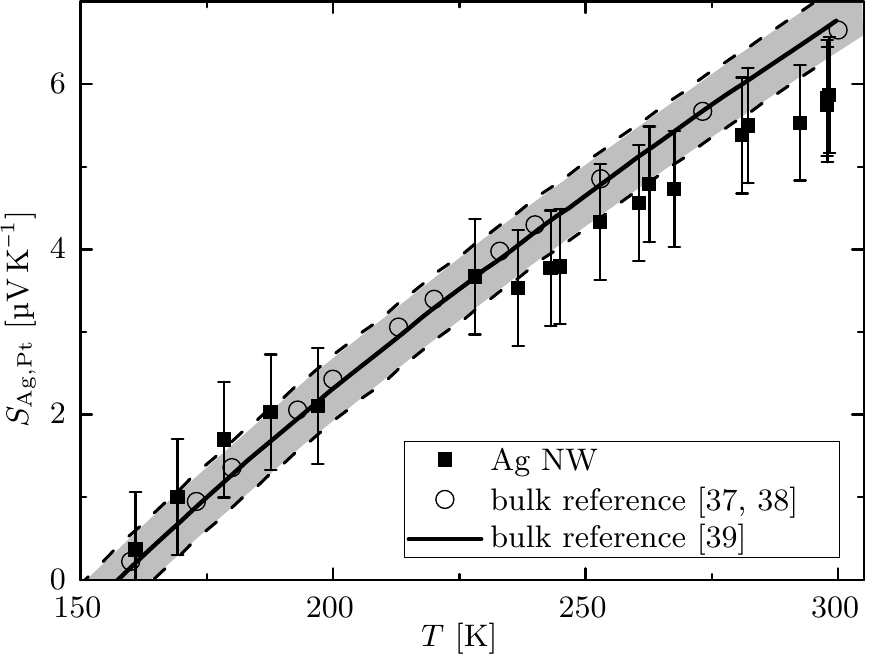}
  \caption{\label{fig:SeebeckRef} Measured Seebeck coefficients $S_{\mathrm{Ag, Pt}}$ of Ag NW1 ($\filledmedsquare$) as function of \TB{} in comparison to bulk values ({\Large $\circ$}).\cite{Schroeder65,Moore73} The solid line shows bulk values from Ref.~\onlinecite{Cusack58}. The shaded area indicates the absolute error of $S_{\mathrm{Ag, Pt}}$ by comparison of the references \onlinecite{Schroeder65,Moore73, Cusack58,Roberts1981,Roberts1977}.}  
\end{figure}

The values of $S_{\mathrm{Ag, Pt}}(T)$ and their temperature dependence are in agreement with the reference values as shown in Fig.~\ref{fig:SeebeckRef}. The reference values were calculated by subtracting the Seebeck coefficients of bulk Ag and Pt samples, which both were measured with respect to lead.\cite{Schroeder65, Moore73} Absolute Seebeck measurements of $S_{\mathrm{Ag}}$ and $S_{\mathrm{Pt}}$ yield the solid line in Fig.~\ref{fig:SeebeckRef}.\cite{Cusack58} The shaded area gives the absolute uncertainty by comparison of the references \onlinecite{Schroeder65,Moore73, Cusack58,Roberts1981,Roberts1977}. 

\subsection{Temperature-dependent Thermal Conductivity and Lorenz Number}
\label{sec:ThermalConductivity}

\begin{figure}[htbp]
  \includegraphics[width=\columnwidth]{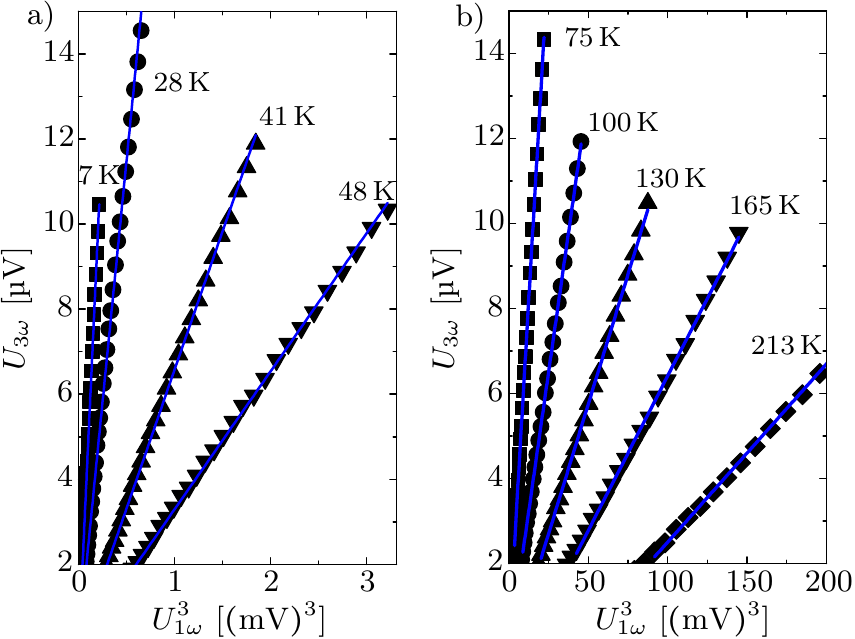}
  \caption{Measured $U_{3\omega}$ of Ag NW4 as function of $U_{1\omega}^3$ for (a) the temperature range from \SI{7}{\kelvin} to \SI{48}{\kelvin} and (b) the temperature range from \SI{75}{\kelvin} to \SI{213}{\kelvin}.}
  \label{fig:U1U3}
\end{figure}

The $3\omega$-method\cite{Cahill1990} was applied in four-terminal sensing geometry in order to determine the thermal conductance $G$ and the thermal conductivity $\lambda$ of the NW. Therefore, the first harmonic of the voltage drop $U_{1\omega}$ and the third harmonic of the voltage drop $U_{3\omega}$ at the NW were measured by applying a sinusoidal AC voltage of the frequency $f$.\cite{Kimling11} A series resistance of \SI{10}{\kilo \ohm} was used to reduce the current through the NW as well as to justify the usage of an voltage source.\cite{Dames05} For the frequency band between \SI{100}{\hertz} and \SI{1}{\kilo \hertz} the voltages $U_{1\omega}$ and $U_{3\omega}$ were measured. Both were constant with respect to frequency as required.\cite{Dames05, Touloukian73} In the vacuum, the measurements of $U_{1\omega}$ and $U_{3\omega}$ were performed at the frequency $f = \SI{123}{\hertz}$ and at different \TB{} in the range between \SI{1.4}{\kelvin} and \RT{}. We found $U_{3\omega}$ to be proportional to $U^3_{1\omega}$ in the whole temperature range as required and exemplarily depicted in Fig.~\ref{fig:U1U3}. 

Due to random fluctuation of $U_{3\omega}$ of about \SI{1}{\micro\volt} at low voltages and for $T > \SI{150}{\kelvin}$ the data analysis was performed for $U_{3\omega} > \SI{2}{\micro\volt}$. For $T < \SI{150}{\kelvin}$ the fluctuations disappeared and all $U_{3\omega}$ were taken into account. The slope of $U_{3\omega}(U^3_{1\omega})$ correlates with the measured thermal conductance $G_\mathrm{th,exp}$ as described by Eq.~\ref{eq:conductance}\cite{Dames05} 
\begin{equation}
  G_\mathrm{th,exp} = \frac{\alpha }{24 R_\mathrm{NW}} \frac{U^3_{1\omega}}{U_{3\omega}}, 
  \label{eq:conductance}
\end{equation}

where $R_\mathrm{NW}$ is the NW's electrical resistance (see Fig.~\ref{fig:Resistance} and Fig.~\ref{fig:Bloch}) and $\alpha$ is the NW's TCR (see inset of Fig.~\ref{fig:Bloch}). Then, the NW's thermal conductivity results from ${\lambda_\mathrm{exp} = G_\mathrm{th,exp} 4 l_\mathrm{NW}/(\pi d^2)}$ which yields a value of \SI{220 \pm 30}{\watt \per \meter \per \kelvin} for NW4 at RT. This value is reduced with respect to the bulk material's value of ${\lambda_{\mathrm{B}} = \SI{429}{\watt \per \meter \per \kelvin}}$.\cite{Ho1972} The reduction of $\lambda$ can be explained by the reduced electrical conductivity within the framework of the Wiedemann-Franz law. 

\begin{figure}[htbp]
  \includegraphics[width=\columnwidth]{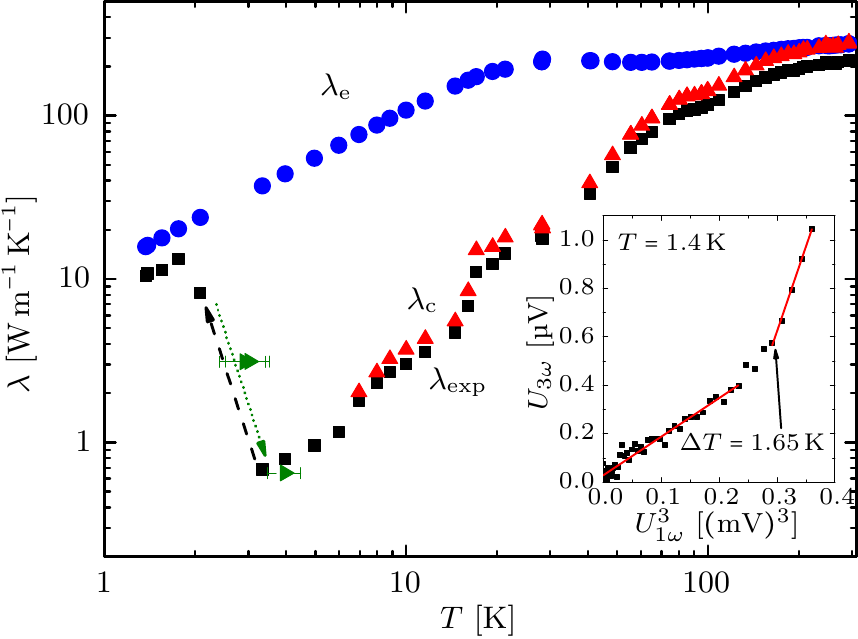}
  \caption{\label{fig:lambda} Measured thermal conductivity $\lambda_\mathrm{exp}$ ($\filledmedsquare$) of Ag NW4 and the calculated thermal conductivity $\lambda_\mathrm{e}$ (\textcolor{blue}{\Large $\bullet$}) from a constant bulk Lorenz number $L_\mathrm{Ag}$. Corrected thermal conductivity $\lambda_\mathrm{c}$ (\textcolor{red}{\Large $\filledtriangleup$}) for a thermal contact resistance arising from the condition $\lambda_c(\RT{}) = \lambda_e(\RT{}) = L_\mathrm{Ag} \sigma(\RT{}) \RT{}$. The dashed arrow marks the return of $\lambda_\mathrm{exp} \rightarrow \lambda_\mathrm{e}$ for $T < \SI{2}{\kelvin}$, whereas the dotted arrow and (\textcolor{OliveGreen}{\Large $\blacktriangleright$}) shows the reverse characteristic of $\lambda$ due to Joule heating of the NW. The inset shows $U_{3\omega}(U_{1\omega}^{3})$ at $T=\SI{1.4}{\kelvin}$. For a mean temperature increase $\Delta T > \SI{1.65}{\kelvin}$ additional values for $\lambda$ are evaluated which are shown by (\textcolor{OliveGreen}{\Large $\blacktriangleright$}).}
\end{figure}

Initially, the Wiedemann-Franz law $\lambda_\mathrm{e}(T) = L \sigma(T) T$ was applied to calculate the electric thermal conductivity $\lambda_\mathrm{e}$
 from the NW's $\sigma(T)$, where $L$ is the temperature-independent Lorenz number. The phonon contribution to the thermal conductivity will be marginal, as a metal is considered here.\cite{Ou2008,Tritt04} At high $T$, the Lorenz number $L$ is expected to have the Sommerfeld value $L_0=(\pi^2/3)(k_B/e)^2=\SI{2.44E-8}{\volt \squared \per \kelvin \squared}$, with the Boltzmann constant $k_B$ and the electron charge $e$. Bulk silver exhibits a reduced Lorenz number of $L_\mathrm{Ag} = 0.96 L_0$ at \RT{}.\cite{Laubitz69} Figure~\ref{fig:lambda} compares the experimentally determined $\lambda_\mathrm{exp}$ with $\lambda_\mathrm{e}$. Whereas $\lambda_\mathrm{exp}$ compares to $\lambda_\mathrm{e}$ at $T \approx \RT{}$, the difference $\lambda_\mathrm{e} - \lambda_\mathrm{exp}$ increases with decreasing $T$. 

For $T < \SI{2}{\kelvin}$ one observes, that $\lambda_\mathrm{exp}$ returns to $\lambda_\mathrm{e}$. As depicted in the inset of Fig.~\ref{fig:lambda}, measurements of $U_{3\omega}(U_{1\omega}^{3})$ at \SI{1.4}{\kelvin} revealed an additional slope at powers above $\SI{0.3}{\micro\watt}$. One observes, that the change of the slope in $U_{3\omega}(U_{1\omega}^{3})$ occurs when the NW's mean temperature $T + \Delta T$ exceeds \SI{3}{\kelvin}. Here, the NW's mean temperature increase $\Delta T$ due to Joule heating is determined by $\Delta T = 2 U_{3\omega}/(\alpha U_{1\omega})$.\cite{Dames05} The additional slopes are evaluated under consideration of $\Delta T$ and yield additional values in $\lambda$. Those values follow the temperature characteristic and match the measured values of $\lambda$ which are determined for $T > \SI{3}{\kelvin}$. Hence, at very low temperatures it is important to determine whether Joule heating influences the observation of $\lambda$.

A weak dip in $\lambda_\mathrm{exp}$ is observed at $T \approx \SI{20}{\kelvin}$. This may give rise to a low temperature peak in thermal conductivity, which generally is observed in high purity bulk materials.\cite{Tritt04} The strong weakening of this feature here accounts to surface scattering. A detailed discussion of $\lambda_\mathrm{exp}$ and a temperature dependent $L(T)$ is given in section \ref{sec:lorenz}.

\begin{figure}[htbp]
  \includegraphics[width=\columnwidth]{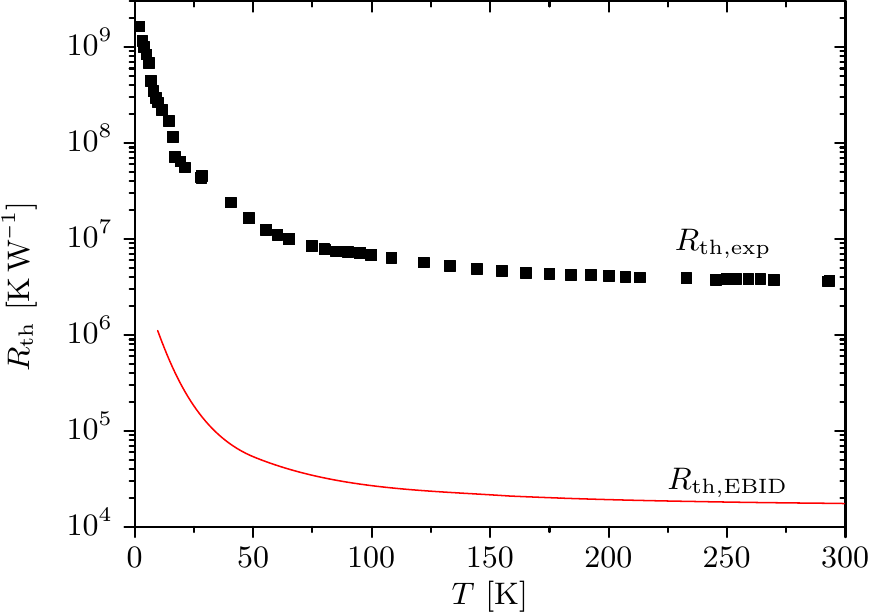}
  \caption{\label{fig:Rth} Measured total thermal resistance $R_\mathrm{th,exp}$ of Ag NW4 compared to the thermal resistance of the EBID contacts $R_\mathrm{th,EBID}$.\cite{Bifano12}}
\end{figure}

In the following we discuss the influence of a thermal contact resistance on $\lambda_\mathrm{exp}$. In order to analyze whether the characteristic of $\lambda_\mathrm{exp}(T)$ originates from a thermal contact resistance, one can compare the experimentally determined thermal resistance ${R_\mathrm{th,exp}(T) = 1/G_\mathrm{th,exp}(T)}$ to the thermal resistance of the contact $R_\mathrm{th, EBID}(T)$. Here, $R_\mathrm{th, EBID}(T)$ can be estimated from the contact area and its thermal conductance. A lower limit of the contact area is given by the contact length and the half NW perimeter and can be calculated to be $\SI{700}{\nano\meter} \times \pi d/2 \approx \SI{1.6E-13}{\meter\squared}$. With the \RT{} value of the thermal conductance of $\SI{170}{\mega \watt\per \meter \squared \per\kelvin}$ for Pt EBID contacts, the total thermal contact resistance of the two inner contacts can be estimated to be ${R_{\mathrm{th,EBID}} = \SI{1.8E4}{\kelvin\per\watt}}$.\cite{Bifano12} At \RT{}, the measured thermal resistance yields $R_\mathrm{th,exp} = \SI{3.6E6}{\kelvin\per\watt}$. Thus, the thermal resistance of the NW dominates the complete thermal resistance. At \RT{}, the thermal resistance of the contacts is lower by two orders of magnitude. Furthermore, $R_\mathrm{th,exp}(T)$ dominates $R_{\mathrm{th,EBID}}(T)$ in the whole temperature range with up to three magnitudes of order at low $T$ as shown in Fig.~\ref{fig:Rth}. Here, $R_{\mathrm{th,EBID}}(T)$ was calculated from the theoretical characteristic of the thermal conductance for EBID contacts.\cite{Bifano12} 

If the thermal contact in Ref. \onlinecite{Bifano12} was underestimated, a correction with respect to higher contact resistances is given by the following approach. Generally, the NW's thermal resistance $R_\mathrm{th, NW}(T)$ is given by $R_\mathrm{th, NW}(T) = R_\mathrm{th,exp}(T) - a R_\mathrm{th,EBID}(T)$ as both thermal resistances are in a serial connection. Here, $a$ is a scalar to vary $R_\mathrm{th,EBID}(T)$. Then, the corrected thermal conductivity is given by Eq. \eqref{eq:lambdacorr}
\begin{equation}
  \lambda_\mathrm{c} = \frac{1}{R_\mathrm{th, NW}} \frac{l}{A}. 
\label{eq:lambdacorr}
\end{equation}

The maximum correction with respect to a thermal contact resistance is given when the Wiedemann-Franz law with $L(\RT{}) = L_\mathrm{Ag}$ is obeyed at \RT{}. This condition yields $a =45$. The corresponding $\lambda_\mathrm{c}$ is given in Fig. \ref{fig:lambda}. One observes, that even the highest possible contact resistance cannot explain the deviation of $\lambda_\mathrm{exp}$ from $\lambda_\mathrm{e}$. Furthermore, the difference $\lambda_\mathrm{c} - \lambda_\mathrm{exp}$ decreases with decreasing $T$. Hence, the Lorenz number has to be considered to be temperature dependent. 

In our experiment $L(T)$ can be determined by Eq. \eqref{eq:LorenznumberI}
\begin{equation}
  L(T) = \frac{\lambda(T)}{\sigma(T) T} = \frac{R_\mathrm{NW}}{R_\mathrm{th,NW}} \frac{1}{T}, 
	\label{eq:LorenznumberI} 
\end{equation}
so that $L(T)$ depends on the NW's electrical and thermal resistance. If the thermal contact resistance can be neglected, $L(T)$ depends on the measured voltage ratios $U_{1\omega}^3(T)/U_{3\omega}(T)$, the TCR $\alpha$ and the bath temperature \TB{} as given by Eq. \eqref{eq:LorenznumberII} 
\begin{equation}
	L(T) =  \frac{\alpha(T)}{24 \TB} \frac{U_{1\omega}^3(T)}{U_{3\omega}(T)}.
	\label{eq:LorenznumberII}
\end{equation}
Thus, $L(T)$ can be determined from the $3\omega$ measurements and is independent of the NW's geometry. The calculated values for $L(T)$ of NW4 were normalized with respect to $L_0$ and are plotted as a function of the reduced temperature $T/\Theta_\mathrm{D,Ag}$ as shown in Fig.~\ref{fig:lorenz}. The detailed discussion on $L(T)$ is given in section~\ref{sec:lorenz}.

\section{Discussion}
\label{sec:discussion}

\subsection{Discussion of Structure and Thermoelectric Measurements}
\label{sec:discussionTEM}
In this study the combined thermoelectrical investigation of individual NWs and the subsequent structural and chemical analysis by TEM methods was applied. Therefore, correlations between the NW's structure and its thermoelectrical properties can be elucidated. The structural analysis revealed a single crystalline growth along the [110] direction. One finds sporadic kinks at which the direction of growth along [110] is preserved and a grain boundary appears. From the thermoelectrical characterized NWs only Ag NW2 showed a pronounced kink. As a consequence of the low grain boundary contribution, grain boundary scattering in the NWs can be neglected. An additional structural feature of our Ag NWs were twin boundaries intersecting the whole NW parallel to the direction of transport. Twin boundaries are known to be weak scatterers. Moreover, the influence of impurities and defects is rather low caused by the high quality of the base materials and the single crystalline structure. Consequently, the transport properties are mainly governed by electron-phonon scattering and/or surface scattering. However, impurity scattering dominates at temperatures below \SI{2}{\kelvin}. 

First of all, the Seebeck coefficients and their temperature dependence are in agreement with those of the bulk material. Generally, the Seebeck coefficient is very sensitive to material compositions, so that the agreement with the bulk values give rise to the material's quality. The electrical conductivity of the Ag NWs showed reduced values with respect to the bulk. Nevertheless, the values of $\sigma(\RT{})$ agree with ensemble measurements on NWs with comparable diameter\cite{Bid2006} and compare with theoretical calculations considering surface scattering.\cite{Huang09} However, at low temperatures we observe deviations to previous experiments and to the classical Bloch-Gr\"uneisen equation. The temperature dependent surface scattering is regarded the dominant contribution because the single crystalline Ag NWs exhibit a low density of defects such as impurities, dislocations, kinks and show no grain boundaries. This high crystal quality is also proved by the $RRR$ of \num{12}, which is high with respect to comparably sized Ag NWs.\cite{Bid2006} Our temperature-dependent measurements confirm, that the reduction in $\sigma(T)$ can be understood in terms of additional scattering at the surface. The increased surface-to-volume ratio of the NW with respect to the bulk promotes surface effects. 

Whereas $\sigma(T)$ showed metallic behavior the temperature dependence of $\lambda_\mathrm{exp}(T)$ was monotonically decreasing with decreasing $T$. Furthermore, $\lambda_\mathrm{exp}(T)$ proved to be less than the bulk values. A similar behavior of the thermal conductivity was observed in $\SI{100}{\nano\meter} \times \SI{180}{\nano\meter}$ nickel  NWs.\cite{Ou2008} Moreover, for the nickel NWs a strong reduction in $L(T)$ is observed for $T<\SI{60}{\kelvin}$. From our measurements we conclude a temperature dependent Lorenz number as well. The temperature dependence of $L(T)$ is considered in the following section.  
 
\subsection{Discussion of the Lorenz Number}
\label{sec:lorenz}

The general characteristic of $L(T)$ in metals is described by Eq. \eqref{eq:LorenzT} given by\cite{Makinson1938,Tritt04}  

\begin{footnotesize}
  \begin{equation}
		\frac{L(T)}{L_0} = \frac{\frac{R_0}{\Reph{}} 
	  + \Big(\frac{T}{\Theta_\mathrm{D}}\Big)^5 J_5\Big(\frac{\Theta_\mathrm{D}}{T}\Big)}
	  {\frac{R_0}{\Reph{}} +
	  \Big(\frac{T}{\Theta_\mathrm{D}}\Big)^5 J_5\Big(\frac{\Theta_\mathrm{D}}{T}\Big) 
	  \bigg[
	  1 + 
	  \frac{3}{\pi^2} \Big(\frac{k_\mathrm{F}}{q_\mathrm{D}}\Big)^2 \Big(\frac{\Theta_\mathrm{D}}{T}\Big)^2 -
	  \frac{1}{2\pi^2}  \frac{J_7\big(\frac{\Theta_\mathrm{D}}{T}\big)}{J_5\big(\frac{\Theta_\mathrm{D}}{T}\big)} 
	  \bigg]}
    \label{eq:LorenzT}
  \end{equation}
\end{footnotesize}

with 
\begin{equation}
  J_n \left(\frac{\Theta}{T} \right) =  \int_{0}^{\frac{\Theta}{T}}{\frac{x^n e^x}{(e^x - 1)^2}} \mathrm{d}x.
  \label{eq:J}
\end{equation}

\begin{figure}[htbp]
  \includegraphics[width=\columnwidth]{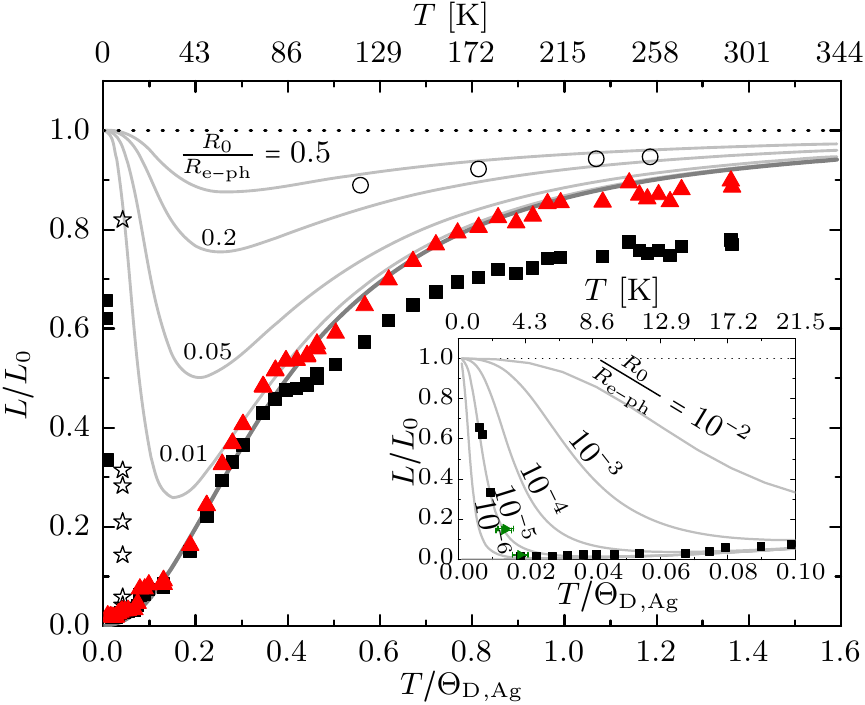}
  \caption{\label{fig:lorenz} Measured Lorenz number $L$ as function of normalized temperature ($\filledmedsquare$) and Lorenz numbers corrected for thermal contact resistance (\textcolor{red}{{\Large $\filledtriangleup$}}) for Ag NW4. Bulk reference values ({\Large $\circ$}) are shown.\cite{White91} Low temperature values of Ag bulk samples with different $RRR$ ($\largestar$).\cite{Gloos90} Measured $L$ of a \ce{Pt} NW of \SI{130}{\nano\meter} diameter ({\large $\triangledown$}).\cite{Volk08} The dark gray solid line shows the Lorenz number of a defect free monovalent metal, whereas bright gray lines represent $L$ for an imperfect metal with different defect concentrations.\cite{Tritt04} The inset shows non-corrected values that compare with the theory of high purity material. Here (\textcolor{OliveGreen}{\Large $\blacktriangleright$}) are determined for the Joule heated NW (compare Fig.~ \ref{fig:lambda}).} 
\end{figure}

Here, $L(T)$ depends on the impurity concentration, which can be expressed by the fraction $R_0/\Reph{}$. The calculated $L(T)/L_0$ for different impurity concentrations are plotted as shown in Fig.~\ref{fig:lorenz} (solid lines). For a non-vanishing $R_0$, $L(T)$ exhibits a minimum and converges to $L_0$ for high temperatures and for $T \rightarrow 0$. The deviation from $L_0$ increases with the material's purity. If $R_0 = 0$, $L(T)$ shows a quadratic behavior for low $T$ and converges to zero for $T \rightarrow 0$. 

We found that the measured $L(T)$ follows the characteristic for pure materials, whereas the return to $L_0$ is found for $T<\SI{2}{\kelvin}$. For $T<0.4\Theta_\mathrm{D,Ag}$ one observes that $L(T)$ is a quadratic function of $T$ in full agreement with the theory. For $T > 0.4 \Theta_\mathrm{D}$, the measured $L(T)$ is reduced with a maximum deviation of \SI{16}{\percent} from the characteristic for pure materials. A similar reduction was previously observed in Pt nanowires of \SI{130}{\nano\meter} diameter.\cite{Volk08} \ce{Ag} compares to \ce{Pt} because both are monovalent metals with a comparable Debye temperature ($\Theta_\mathrm{D,Pt} = \SI{234 \pm 6}{\kelvin}$).\cite{Harris1965} Here, the measured Lorenz numbers for NWs of both materials lie outside the theoretical scope of a metal. However, the  analytical investigation of $L$ in a 2-dimensional system predicts a reduced $L$ at \RT{} and a sharp reduction for very low $T$.\cite{Tripathi2010} The reduction with respect to $L_0$ may also be theoretically explained by consideration of a more appropriate electron-phonon interaction with the real phonon dispersion. 

Here, we analyzed the deviation of $L(T)$ with respect to a thermal contact resistance as follows. Analogous to the calculation of the corrected thermal conductivity $\lambda_\mathrm{c}$ (see section \ref{sec:ThermalConductivity}), a thermal contact resistance $a R_\mathrm{th, EBID}(T)$ was considered and a corrected Lorenz Number was calculated by Eq. \eqref{eq:LorenznumberI}. Generally, assuming a thermal contact resistance leads to a raise in $L$. For $a = 30$, the corrected values of $L(T)$ values match the calculated $L(T)$ for a pure metal in the whole temperature range as shown in Fig.~\ref{fig:lorenz}~({\Large $\filledtriangleup$}). Therefore, the corrected thermal conductivity at \RT{} yields \SI{256}{\watt\per\meter\per\kelvin}, which implies a correction by \SI{16}{\percent} with respect to the non-corrected value. For decreasing $T$ the influence of the thermal resistance decreases, so that the correction has no significant influence on $L(T)$ for $T < 0.4\Theta_\mathrm{D,Ag}$ (compare Fig.~\ref{fig:lorenz}). 

In Fig. \ref{fig:lorenz}, the inset shows that the return of $L(T)$ to $L_0$ was experimentally observed for $T/\Theta_\mathrm{D} < 0.01$. This corresponds to a high purity bulk metal for which $R_0/\Reph{} \approx \num{1E-5}$. In order to rank the NW's quality with respect to Ag bulk, we compared the NW's $L(\SI{9}{\kelvin})/L_0 = 0.02$ with that of high purity Ag bulk samples with $RRR=\num{3300}$.\cite{Gloos90} The Ag bulk samples showed a normalized Lorenz number of $L(\SI{9}{\kelvin})/L_0 = 0.04$, which is higher by a factor of two with respect the NW's $L$. From this point of view, the NW's purity is even higher than that of the high purity bulk Ag.\cite{Gloos90} As previously discussed, the reduced $RRR$ by surface scattering in NWs does not reflect the material's quality with respect to the bulk. Instead, the temperature dependence of $L(T)$ gives the intrinsic properties of the NW. Therefore, the Lorenz number in NWs was found to be independent of surface scattering.

One has to note, that in Fig.~\ref{fig:lorenz} the literature value $\Theta_\mathrm{D, Ag}=\SI{215}{\kelvin}$ was taken for temperature normalization. A lower $\Theta_\mathrm{D}$ as supposed in Ref. \onlinecite{Bid2006, Kamalakar2009} would shift the $L(T)$ values of the entire temperature range outside the possible range. The usage of higher $\Theta_\mathrm{D}$ would move the values into the possible range, whereas none of the theoretical characteristics would be matched. Hence, measurements of $\sigma(T)$ and $L(T)$ reveal an unchanged Debye temperate in Ag NWs. 

\subsection{Discussion of the Electrical and Thermal Mean Free Path}

\begin{figure}[htbp]
  \includegraphics[width=\columnwidth]{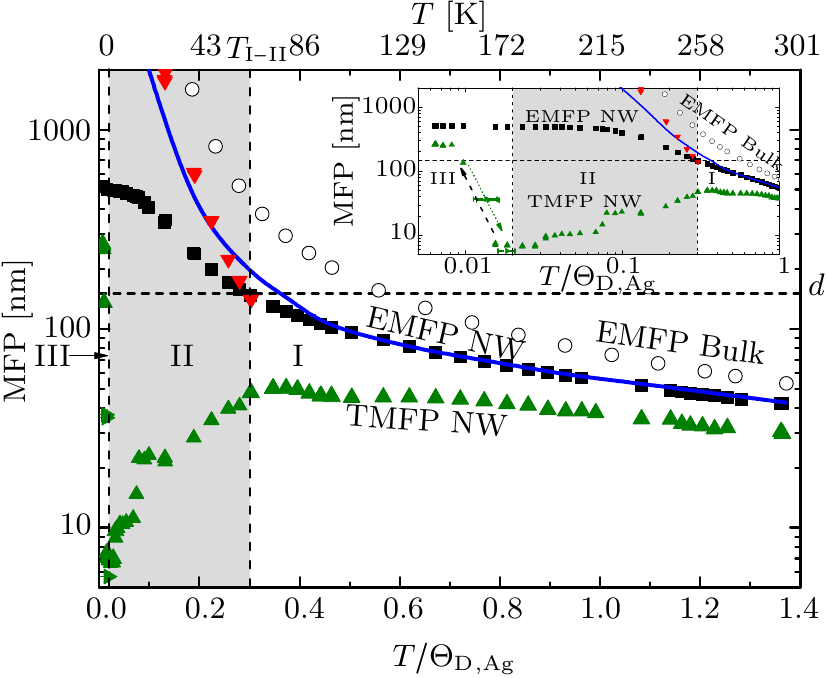}
  \caption{\label{fig:MFP} The determined electron mean free path (EMFP: $\filledmedsquare$) and the thermal mean free path (TMFP: \textcolor{OliveGreen}{\Large $\filledtriangleup, \blacktriangleright$}) of Ag NW4 as function of normalized temperature $T/\Theta_\mathrm{D,Ag}$. The bulk EMFP $\Lambdael{,B}$ is calculated by Eq. \eqref{eq:LambdasI} and depicted as~({\Large $\circ$}). The solid line shows the calculated EMFP from $\Lambdael{,B}(T) \alpha_\mathrm{NW}(T)/\alpha_\mathrm{B}(T)$ which compares with \Lambdael{,NW} for ${T > \Tc{}}$.\cite{White91, Stojanovic10} For $T < \Tc{}$ the determined EMFP along the direction of transport \Lambdael{,d} is shown by (\textcolor{red}{\Large $\filledtriangledown$}). The three regions \RM{1} (diffusive transport), \RM{2} (quasi ballistic transport) and \RM{3} (dominating impurity scattering) are marked. In the inset the low temperature region is shown in logarithmic scale.} 
\end{figure}

In the following, the scattering processes in Ag NWs are discussed on the basis of the temperature dependent electrical mean free path (EMFP) $\Lambdael{,NW}(T)$ and the thermal mean free path (TMFP) $\Lambdath{}(T)$. Here, $\Lambdael{,NW}(T)$ was determined by Eq. \eqref{eq:LambdasII}. The TMFP $\Lambdath{}$ in the NW can be calculated by 
\begin{equation}
  \Lambdath{}(T) = \frac{3 \lambda}{C_e(T) v_\mathrm{F}},
  \label{eq:Lambdath}
\end{equation}
where $C_e(T)$ is the temperature-dependent specific heat of the electron gas and $v_\mathrm{F}$ is the Fermi velocity of Ag. For the calculation an isotropic Fermi sphere was assumed which yields an isotropic $v_\mathrm{F} = \sqrt{2 E_\mathrm{f}/m_\mathrm{e}}$, whereas $C_e(T) = \pi^2 k_\mathrm{B}^2 n T/2E_\mathrm{f}$, with $m_\mathrm{e}$ the electron mass, $n$ the electron density and $E_\mathrm{f}$ the Fermi energy.\cite{Fisher2013} The EMFP and TMFP are plotted as function of normalized temperature $\TB/\Theta_\mathrm{D, Ag}$ as shown in Fig.~\ref{fig:MFP}. 

From the TEM analysis, we concluded that the transport properties in Ag NWs will not depend on grain boundary scattering or scattering on twins. Hence, electron-phonon scattering, electron-surface scattering and impurity scattering can determine the transport properties. We defined the three regions \RM{1}, \RM{2} and \RM{3} (see Fig.~\ref{fig:MFP}) to discuss the influence of the scattering mechanisms on the NWs' transport characteristics.

Beginning at \RT{} (region \RM{1}), the EMFP and TMFP of the NW are increasing with decreasing temperature. Here, the increase of the TMFP is less than in the EMFP. Down to $\Tc{} \approx 0.3 \Theta_\mathrm{D,Ag}$ one observes diffusive scattering where electron-phonon scattering is dominant and the probability of large angle scattering events decreases with decreasing temperature.\cite{Tritt04} Here, every scattering event can cause backscattering that give rise of an electrical and thermal resistance. Furthermore, scattering events on the surface influence the charge and heat transfer. Thus, EMFP and TMFP show reduced values with respect to the bulk EMFP. The scattering length \Lambdael{,SS} for surface events was determined by the Matthiessen rule. The \Lambdael{,SS} are in the order of the NWs' diameters and give evidence of surface scattering as well. The measured TCR $\alpha_\mathrm{NW}$ is reduced with respect to the bulk material, which is due to surface scattering. Scaling the bulk material's EMFP with the ratio $\alpha_\mathrm{NW}(T)/\alpha_\mathrm{B}(T)$ the resulting graph follows the NW's EMFP for $T > \Tc{}$ as shown in Fig.~\ref{fig:MFP}. This result confirms the assumption that the reduced value of the TCR is due to a reduced $\Lambdael{,NW}$ and therefore can be used as a measure of the scattering length along the transport direction. 

At $T = \Tc{}$ the EMFP compares with the NW diameter. Estimating the scattering angle $\gamma$ by $\gamma = \arcsin(\frac{2.82 k_\mathrm{B} T}{2 v_\mathrm{s} \hbar k_\mathrm{F}})$ one finds an average scattering angle of $\gamma(\Tc{}) \approx \SI{25}{\degree}$.\cite{Tritt04, Fisher2013} For this scattering angle the EMFP is $\Lambdael{,NW} = \frac{d}{2 \sin \gamma} \approx \SI{200}{\micro \meter}$ which compares to the experimental value $\Lambdael{,NW}(\Tc{})$ and $\Lambdael{,SS}$. 

For $0.01 \Theta_\mathrm{D,Ag} < T < \Tc{}$ (region \RM{2}) one finds $\Lambdael{} > d$ and therefore quasi ballistic electron transport within the NW. Here, small angle electron scattering at phonons appears, whereas the scattering angle decreases with decreasing $T$. Due to the  low value of $\gamma$ many events are required before one random large angle process can effect an increase in electrical resistance. Furthermore, low angle scattering is an inelastic process, so that every scattering event can increase the thermal resistance. Due to surface scattering, the EMFP increases less than in bulk material but the ratio $\Lambdael{,NW}/\Lambdath{} \propto L(T) \propto T^2$ is conserved. Therefore, the TMFP significantly starts to decrease with decreasing temperature at $\Tc{}$. Simultaneously, the influence of the surface on the EMFP begins to decrease as described by Eq. \eqref{eq:BlochX}, so that the EMFP additionally is increasing. For $T < 0.1\Theta_\mathrm{D,Ag}$ the EMFP starts to saturate. 

One has to remark that the determined EMFP in NWs is given by the integral of the EMFPs over all directions. As the transport along the vertical direction is limited by the NW's diameter the EMFP along the direction of transport can be even longer. Using a rough estimation, the experimental value \Lambdael{,NW} follows from the geometric mean $\Lambdael{,NW} = \sqrt[3]{d^2 \Lambdael{,d}}$ of the transversal MFP limited by the diameter $d$ and the longitudinal MFP $\Lambdael{,d}$ as considered to be along the direction of transport. For $0.1 \Theta_\mathrm{D,Ag} < T < 0.3 \Theta_\mathrm{D,Ag}$ we observe that $\Lambdael{,d}$ compares well with the bulk EMFP scaled by $\alpha_\mathrm{NW}(T)/\alpha_\mathrm{B}(T)$, the ratio of the TCRs, as shown in Fig.~\ref{fig:MFP}. Hence, the NW's EMFP in region~\RM{1} and mostly in region~\RM{2} can be approximated by $\Lambdael{,B}$ and the TCRs' ratio. 

If $T < 0.01 \Theta_\mathrm{D}$ (region \RM{3}) the EMFP is large enough to exceed the distance of impurities. Then, scattering at impurities realizes larger scattering angles again, so that the TMFP increases and $L(T)$ returns to $L_0$. Using the estimation above we find $\Lambdael{,d}(\SI{1.4}{\kelvin}) \approx \SI{6}{\micro\meter} > \Lambdael{,NW}(\SI{1.4}{\kelvin}) \approx \SI{0.5}{\micro\meter} \gg d = \SI{0.15}{\micro\meter}$. Hence, the transmission $\phi = \Lambdael{}/(\Lambdael{}+l_\mathrm{NW})$ yields \SI{3}{\percent} for \Lambdael{,NW} and \SI{30}{\percent} for \Lambdael{,d} at $T=\SI{1.4}{\kelvin}$.\cite{Fisher2013} 

\subsection{Thermoelectric Figure of Merit}

By the measurement of $S_\mathrm{Ag,Pt}$ and $L$ the thermoelectric figure of merit of the Ag NWs with respect to Pt contacts can be calculated by Eq.~\eqref{eq:ZT} and yield $ZT = \num{2E-3}$ at RT. 
\begin{equation}
  ZT = \frac{S_\mathrm{Ag,Pt}^2 \sigma}{\lambda} \TB{} = \frac{S_\mathrm{Ag,Pt}^2}{L} 
\label{eq:ZT}
\end{equation} 
The low $ZT$ value is reasonable due to the low Seebeck coefficient, which is expected for a pure metallic material. With decreasing $T$ the thermoelectric figure of merit decreases. At $T \approx \SI{150}{\kelvin}$ a minimum with $ZT = 0$ is expected as $S_\mathrm{Ag,Pt}(T)$ exhibits a change in sign.\cite{Schroeder65,Moore73} Below \SI{150}{\kelvin} the value of $ZT$ will increase again as the absolute value of $S_\mathrm{Ag,Pt}$ increases. For further lowering of $T$, $ZT(T)$ is expected to run through a maximum as $L(T) > 0$ and $S_\mathrm{Ag,Pt}(T) = 0$ for $T \rightarrow 0$. 

\section{Conclusion}
In this study the temperature-dependent thermoelectric properties --- electrical conductivity $\sigma$, thermal conductivity $\lambda$ and Seebeck coefficient $S$ --- as well as the structural and chemical composition of individual highly pure single crystalline metallic Ag NWs were measured. This comprehensive investigation of individual NWs allows to draw conclusions from the NWs structure to the thermoelectric properties. The temperature-depended thermopower was found to be in agreement with the bulk. In contrast, a reduction in $\sigma(T)$ and $\lambda(T)$ with respect to the bulk was found and can be attributed to surface scattering. A modified Bloch-Gr\"uneisen formula was applied to match the low temperature resistances, where the transition from diffusive to quasi ballistic transport has to be taken into account. 

The Lorenz number $L(T)$ in NWs was found to be independent of surface scattering and instead reflects the material's purity. Especially at low $T$, the Ag NW's $L(T)$ follows the theory of a high purity material.\cite{Tritt04} However, in NWs the material's quality can not be concluded from the residual resistance ratio $RRR$ as for the bulk, because a reduction in the $RRR$ may be a consequence of size-dependent surface scattering. 

At high $T$, $L(T)$ for the Ag NWs was found to be reduced with respect to the bulk, as previously found in Pt NWs.\cite{Volk08} Here, the correction with respect to an increased thermal contact-resistance $L(T)$ leads to the agreement with the theory in the full temperature range. However, a reduced Debye temperature $\Theta_\mathrm{D}$ for NWs, as proposed in the literature\cite{Bid2006,Kamalakar2009}, could not be confirmed. A deviation from the bulk $\Theta_\mathrm{D,Ag}$ would give an inconsistent description of $\sigma(T)$ and $L(T)$.

The comprehensive investigation as presented here is transferable to a multitude of material systems. In particular, optimization of $ZT$ in semiconducting NWs is a topical issue for which the combined thermoelectrical and structural analysis are prerequisite to study correlations between composition and properties.\cite{Kojda2014} 

\begin{acknowledgments}
The authors thank Prof. Dr. K. Rademann for scientific discussions. This work was funded by the DFG within the priority program SPP1386.
\end{acknowledgments}

\bibliography{References}

\end{document}